\documentclass[11pt]{article}
\pdfoutput=1
\usepackage{graphicx,epsfig}
\usepackage{color}
\usepackage[colorlinks,citecolor=red,linkcolor=blue]{hyperref}
\usepackage{slashed}

\usepackage{epsfig,float,amsmath,amsfonts,amssymb,graphicx,bm,bbm,array,dcolumn,mathabx}
\usepackage{subfigure}

\usepackage[left=2.5cm, right=2.5cm, top=2cm]{geometry}
\usepackage{amsmath, amssymb, mathrsfs, graphics, setspace}
\newcommand{\mathsym}[1]{{}}
\newcommand{\unicode}[1]{{}}

\newcommand{\half}{\frac{1}{2}}

\def\beq{\begin{equation}}
\def\eeq{\end{equation}}
\def\bea{\begin{eqnarray}}
\def\eea{\end{eqnarray}}
\def\bmat{\begin{pmatrix}}
\def\emat{\end{pmatrix}}
\newcommand{ \slashchar }[1]{\setbox0=\hbox{$#1$}   
   \dimen0=\wd0                                     
   \setbox1=\hbox{/} \dimen1=\wd1                   
   \ifdim\dimen0>\dimen1                            
      \rlap{\hbox to \dimen0{\hfil/\hfil}}          
      #1                                            
   \else                                            
      \rlap{\hbox to \dimen1{\hfil$#1$\hfil}}       
      /                                             
   \fi}                                             %

\def\to{\rightarrow}
\def\Chi{{\cal X}}
\def\Dp{{D}}
\def\Up{{U}}
\def\LInt{{\cal L}_{\rm int}}
\def\LIntSS{{\cal L}^{\rm SS}_{\rm int}}
\def\LIntVV{{\cal L}^{\rm VV}_{\rm int}}

\def\LDirMass{{\cal L}_{\rm Dirac\, mass}}
\def\LMajMass{{\cal L}_{\rm Maj\, mass}}
\def\LMass{{\cal L}_{\rm mass}}
\def\Umat{{\cal U}}
\def\aldot{{\dot\alpha}}
\def\bedot{{\dot\beta}}
\def\hc{{\rm h.c.}}

\def\AsymB{{\cal A}_B}
\def\AsymBsig{{\cal A}_B^{(\sigma)}}
\def\AsymBsign{{{\cal A}_B^{n(\sigma)}}}
\def\AsymBsigHat{\hat{\cal A}_B^{(\sigma)}}
\def\ampA{{\cal A}}
\def\ampAc{{\cal A}^c}

\def\GtVmu{\widetilde{G}_V^\mu}
\def\GtVmuc{\widetilde{G}_V^{\mu\,c}}
\def\GtVmuSt{\widetilde{G}_V^{\mu\,*}}

\def\GtVBmu{{\widebar{\widetilde{G}}_V^\mu}}
\def\GtVBmuc{{\widebar{\widetilde{G}}_V^{\mu\,c}}}
\def\GtVBmuSt{{\widebar{\widetilde{G}}_V^{\mu\,*}}}
\def\GVn{{G_V^n}}
\def\ghLn{\hat{g}_{L_n}}

\def\ghRn{\hat{g}_{R_n}}

\def\GZn{{G_\Lambda^n}}

\def\GVBn{{\phantom{\,}\bar{G}_V^n\phantom{\,\!}}}

\def\GZBn{{\phantom{\,}\bar{G}_\Lambda^n\phantom{\,\!}}}

\def\bra#1{\left< #1 | \right.}

\def\matel#1#2#3{\left< #2| #1 | #3 \right>}

\def\nnbar{$n\!-\!\bar{n}$\ }
\def\NNbar{$N\!-\!\bar{N}$\ }

\begin{document}

\title{\normalsize{\bf Effective Theory for Baryogenesis with a Majorana Fermion Pair Coupled to Quarks}}

\author{\normalsize{Shrihari Gopalakrishna$^{a,b}$\thanks{shri@imsc.res.in}\ , Rakesh Tibrewala$^c$\thanks{rtibs@lnmiit.ac.in}}
\\
$^a$~\small{Institute of Mathematical Sciences (IMSc), Chennai 600113, India.}\\
$^b$~\small{Homi Bhabha National Institute (HBNI), Anushaktinagar, Mumbai 400094, India.}\\
$^c$~\small{The LNM Institute of Information Technology (LNMIIT), Jaipur 302031, India.}
}

\maketitle

\begin{abstract}

  With a goal toward explaining the observed baryon asymmetry of the Universe,
  we extend the standard model (SM) by adding a vector-vector dimension-six effective operator
  coupling a new Dirac fermion $\chi$,
  uncharged under the SM gauge symmetries but charged under baryon number,
  to a quarklike up-type fermion and two identical down-type fermions.
  We introduce baryon number violation by adding Majorana masses to $\chi$,
  which splits the Dirac fermion into two Majorana fermions with unequal masses.
  We speculate on the origin of the effective operator, the Majorana mass,
  and the new physics sector connection to the SM,
  by considering some ultraviolet completion examples.
  In addition to the baryon number violation,
  we show that $C$ and $CP$ invariances can be violated in the theory,
  and the interference between tree and loop amplitudes with on-shell intermediate states
  can lead to a baryon asymmetry in $\chi$ decay and scattering processes.
  We write down the Boltzmann equation for baryon number in the early Universe incorporating the decay and scattering baryon asymmetries.
  We provide numerical estimates for the baryon asymmetry generated,  
  and for the neutron-antineutron oscillation rate. 
  
\end{abstract}




\section{Introduction}
\label{Intro.SEC}

It has become established observationally over the last few decades that there is an excess of baryons over antibaryons in the Universe,
the so-called baryon asymmetry of the Universe (BAU).
From a particle physics perspective, although many proposals exist, the mechanism by which the BAU has arisen is yet to be settled.
Irrespective of the specific mechanism, the generation of the BAU requires that the three Sakharov conditions~\cite{Sakharov:1967dj} be satisfied in any candidate theory,
namely,
(i) $C$ and $CP$ violation, (ii) baryon number violation, and, (iii) departure from thermal equilibrium.
Although the standard model (SM) of particle physics has the necessary ingredients,
with the baryon number violation provided by nonperturbative effects (instantons~\cite{tHooft:1976rip} or sphalerons~\cite{Klinkhamer:1984di}),
it is commonly held that it cannot generate the observed BAU,
falling short by not having a big enough $CP$ violating effect,
and, for the now measured Higgs mass, by not having enough departure from equilibrium when the electroweak sphalerons are operative.
This perhaps suggests that some beyond the standard model (BSM) physics is responsible for generating the BAU. 
Many BSM proposals for generating the BAU exist~(for reviews, see for example Refs.~\cite{Kolb:1990vq,Cline:2006ts}),
including electroweak (EW) baryogenesis, leptogenesis, grand unified theory (GUT) baryogenesis, etc.,  
but it is not known which of these, if any, really generated the observed BAU. 

With an aim to generate the required baryon asymmetry, in this work we extend the SM
by introducing a new BSM Dirac fermion $\chi$, uncharged under the SM gauge symmetries,
coupled to an up-type ($\Up$) and two down-type ($\Dp$) quarklike fermions.
This interaction assigns nonzero baryon number to $\chi$.
We introduce baryon number violation by introducing a Majorana mass for $\chi$, which splits the $\chi$ into two Majorana fermions $\Chi_n$ with unequal masses. 
We write down a BSM effective theory which has a rich enough structure to satisfy the Sakharov conditions and generate the BAU {\em directly}
by the decay and scattering processes involving the $\Chi$.
We consider an effective theory with a dimension-six four-fermion operator of the
$(\Chi\Up)(\Dp\Dp)$ type, where the parentheses indicate which spinors are Lorentz contracted,
and we analyze in detail the scalar-scalar (SS) and vector-vector (VV) Lorentz structures in the interactions. 
Our work thus belongs to the so-called ``neutron-portal'' class of models since this operator couples the
electromagnetic (EM) charge neutral $\Chi$ to the $UDD$ operator that has quantum numbers of the neutron.
The Sakharov conditions are satisfied in our proposal, since, 
(i) phases in the interaction of $\Chi$ to quarks and in the $\Chi$ Majorana mass lead to $C$ and $CP$ violation,
(ii) the interaction assigning nontrivial baryon number to $\Chi$ implies that the $\Chi$ Majorana mass gives baryon number violation,
and,
(iii) the decay and scattering processes of the $\Chi$ in the backdrop of the Hubble expansion of the Universe leads to a departure from thermal equilibrium.
We do not rely on sphaleron dynamics to generate the baryon asymmetry as in electroweak baryogenesis~\cite{Kuzmin:1985mm},
but rather directly generate a quark-level asymmetry.
Our mechanism is in the spirit of leptogenesis~\cite{Fukugita:1986hr} in which a lepton asymmetry is directly generated,
but it differs in that we directly generate a baryon asymmetry.

In our effective theory, proton decay does not occur at the perturbative level since lepton number remains a good classical symmetry of our
Lagrangian density,
while only baryon number is broken, and for proton decay to occur, both these symmetries must be broken.
$\Chi$ exchange induces the baryon number violating neutron-antineutron (\nnbar\!) oscillation,
which is being searched for in experiments.
If the $\Chi$ mass is low enough, a positive signal could result,
while if it is too heavy, null results will place nontrivial constraints on the new-physics parameter space. 

Next, we contrast our work with other works in the literature that have similar ingredients to ours,
namely, the generation of the baryon asymmetry via dimension-six operators involving the $\Chi$ and the $\Up,\Dp$. 
An important distinction between our work and the other earlier works is that  
others consider higher-dimensional scalar-scalar (SS) interaction, 
while we focus on the vector-vector (VV) interaction.
We show in this work that by suitably antisymmetrizing in the color indices,
the VV interaction allows the $(\Chi \Up) (\Dp \Dp)$ interaction with two {\em same} flavor $\Dp$ fields,
while this is forbidden for the SS interaction.
Another distinction is that we focus on the baryon asymmetry generation through one effective operator,
namely, the $(\Chi\Up)(\Dp\Dp)$ VV interaction as mentioned above, 
while other works involve contributions with two different operators being present.
We include both decay and scattering processes in computing the baryon asymmetry, while many other works only include decay. 

Some of the other works that have substantial overlap with our work generate the baryon asymmetry as follows:
Ref.~\cite{Yanagida:1980cy} in decays with the SS interaction in a GUT framework
with the $(\Chi \Dp) (\Chi \Dp)$ and the $(\Chi \Dp) (\Up \Dp)$ operators
present; 
Ref.~\cite{Cheung:2013hza} in decays with the $(\Chi \Up) (\Dp_1 \Dp_2)$ and $(\Chi \Up) (\Up \Chi)$ SS interactions 
necessarily involving two {\em different} $\Dp_{1,2}$ flavors; 
Ref.~\cite{Grojean:2018fus} in decay and scattering processes involving the
$(\Chi \Dp) (\Up \Dp)$ and $(\Chi \Up) (\Up \Chi)$ SS interactions with a single $\Dp$ flavor;
Ref.~\cite{Farrar:2005zd} in $\Chi$ scattering on SM quarks with a specific SS operator with two SU(2) doublet quarks and one singlet;
and,
Ref.~\cite{Baldes:2014rda} in scattering channels with a general set of SS operators.
Other works with a somewhat different focus for the generation of baryon asymmetry include, 
Refs.~\cite{Babu:2006xc,Babu:2008rq,Patra:2014goa} which consider the decay rate asymmetry of a baryon number violating scalar into six quarks vs. antiquarks;
Ref.~\cite{Aitken:2017wie,Ghalsasi:2015mxa} via the oscillation of baryons into antibaryons;
Refs.~\cite{Elor:2020tkc,Elahi:2021jia} through SM $CP$ violation in charm and bottom meson decays and in conjunction with a dark sector;  
in supersymmetric extensions in Refs.~\cite{Babu:2006wz,Allahverdi:2010im,Gu:2011fp,Cui:2013bta,Arcadi:2015ffa,Calibbi:2017rab};
and,
in theories with new colored scalars in Refs.~\cite{Gu:2011ff,Babu:2012vc,Arnold:2012sd,Babu:2013yca,Allahverdi:2013mza,Herrmann:2014fha,Allahverdi:2017edd}. 
For a more comprehensive list of related references, we refer the reader to Ref.~\cite{Grojean:2018fus}.   

This paper is organized as follows.
In Sec.~\ref{Th.SEC} we lay down the effective theory with the $\chi$ and $\Up,\Dp$.
In Sec.~\ref{UVcompl.SEC} we speculate on the origin of our effective theory from some ultraviolet (UV) completion possibilities,
ways in which the baryon number violating sector is coupled to the SM.
The simplest UV completion we write takes the $\Up,\Dp$ as new BSM quarklike vectorlike fermions, and the subsequent analysis primarily focuses on this.
However, we also present UV completion examples where the $\Up,\Dp$ could be identified with SM chiral quarks,
but the phenomenological implications of this possibility is somewhat cursory in this work. 
In Sec.~\ref{ABgen.SEC} we discuss in general terms how in our theory the interference of tree and loop amplitudes yields a baryon asymmetry,
considering the baryon asymmetry generation in the $\Chi_n$ decay process in Sec.~\ref{ABdec.SEC},
and in $\Chi_n$ scattering processes in Sec.~\ref{ABscat.SEC},
and provide numerical estimates of the size of the baryon asymmetry that could be generated.
In a follow-up work~\cite{OurBGChiFeynNum.BIB}, we give a concrete implementation of the baryon number generation mechanism discussed here  
through specific Feynman diagrams at tree and loop levels,
and compute more accurately the baryon number asymmetry that is generated in our effective theory.
In Sec.~\ref{BAUgen.SEC} we write the Boltzmann equations for the $\Chi$ number density and for baryon number density in the early Universe
and obtain an estimate for the $M_\chi$ mass scale indicated if $\Chi$ is to deviate from equilibrium in the expanding Universe.
In a follow-up work~\cite{OurBGChiCosmo.BIB}, we numerically solve the Boltzmann equations and obtain a more accurate computation of the
BAU obtained in our theory. 
In Sec.~\ref{DelB2.SEC} we consider operators that violate baryon number by two units ($\Delta B = 2$) and estimate the \nnbar oscillation rate.
We offer our conclusions in Sec.~\ref{Concl.SEC}.
In Appendix~\ref{spinorAlg.SEC} we provide a compilation of some basic spinor algebra details that we find useful,
and we show there that the SS interaction $(\Chi \Up) (\Dp \Dp)$ with two same flavor $\Dp$ fields is forbidden.
We give details on the diagonalization of the $\Chi$ sector in Appendix~\ref{AppChiM.SEC}.

\section{The Effective Theory}
\label{Th.SEC}

We add to the SM a new Dirac fermion $\chi$ (i.e. a vectorlike fermion)
that is electromagnetic (EM) charge neutral ($Q=0$),
and couple it to 
an up-type color-triplet quarklike fermion $\Up$ with EM charge $Q(\Up) = +2/3$, 
and to color-triplet down-type quarklike fermions $\Dp$ with EM charge $Q(\Dp) = -1/3$.
For the $\chi$, in addition to the Dirac mass, 
we introduce Majorana masses also, which splits the Dirac fermion into two Majorana fermions $\Chi_n$, 
with $n=1,2$.
In this section, we write a general effective Lagrangian with these fields, 
without specifying at this stage which UV completion generates this interaction,
and discuss some UV completion examples in Sec.~\ref{UVcompl.SEC}. There we also identify other related effective operators.
We assume that the usual SM Lagrangian density is present and will not explicitly write it down here,
but will only show the BSM additions. 

We write a general effective interaction including scalar, pseudoscalar, vector and pseudovector Lorentz structures in the couplings as 
\bea
\LInt &=& \frac{1}{\Lambda^2} \left[ \overline{\Dp^c_b} \, \Gamma_{\rm Lor} \left(\frac{\tilde{g}_L}{2} P_L + \frac{\tilde{g}_R}{2} P_R \right) \Dp_a \right]
\ \left[  \bar{\chi} \, \Gamma_{\rm Lor} \left( g_L P_L + g_R P_R \right) \Up_c \right]  \, \epsilon^{abc} + h.c. \ ,
\nonumber \\
         &\equiv&  \frac{1}{\Lambda^2} \left[ \overline{\Dp^c_b} \, \left(\frac{\widetilde\Gamma}{2}\right) \Dp_a \right]
\ \left[  \bar{\chi} \, \Gamma  \Up_c \right]  \, \epsilon^{abc} + h.c. \ ,
    \label{LeffIntA.EQ}
\eea
where
the Lorentz structure $\Gamma_{\rm Lor} \otimes \Gamma_{\rm Lor}$ is $1\otimes 1$ for the scalar-scalar ($SS$) interaction,
and is $\gamma^\mu \otimes \gamma_\mu$ for the vector-vector ($VV$) interaction,
$\Lambda$ is the scale of even heavier new physics that has been integrated out.
To allow for the possibility of $C$ and $CP$ violation in our theory, we take the $\tilde{g}_{L,R}$ and $g_{L,R}$ to be complex. 
The charge conjugated 4-component spinor $\psi^c$ (as defined in Appendix~\ref{spinorAlg.SEC}) has all charges reversed with respect to $\psi$.
As already mentioned, the superscript $c$ on a 4-spinor stands for the charge conjugate spinor,
while the subscript $c$ is a color index that is contracted with the $\epsilon$ tensor.
We form a color singlet antisymmetric combination of the SU(3) fundamental color indices $\{a,b,c\}$. 
$\overline{\Dp^c}$ transforms as $\Dp$ under Lorentz and internal symmetries, and therefore $\LInt$ involves two identical $\Dp$ fermion fields.\footnote{
  One could consider an effective operator with two different EM charge $-1/3$ fields, say, $\Dp_1$ and $\Dp_2$,
  but for the sake of minimality we mostly work with this operator with the identical $\Dp$ fields in our work. 
}
With this effective interaction added,
the $U(1)_B$ baryon number symmetry continues to be a good symmetry at the classical level as in the SM, 
and we can assign baryon number charges $B(\chi)=+1$, and, $B(\Up) = B(\Dp) = +1/3$ as for SM quarks.

Explicitly writing out the Hermitian conjugate terms,
and writing the Lagrangian equivalently using conjugate fields, we write two equivalent forms of Eq.~(\ref{LeffIntA.EQ}) that are useful to us, namely
\bea
\LInt &=& \frac{1}{2\Lambda^2} \left\{
 \left[ \overline{\Dp^c_b} \, \widetilde\Gamma \Dp_a \right] \ \left[  \bar{\chi} \, \Gamma \Up_c \right] +
 \left[ \overline{\Dp}_a \, \widebar{\widetilde\Gamma} \Dp^c_b \right] \ \left[ \bar{\Up}_c \, \widebar\Gamma \chi \right]
 \right\} \, \epsilon^{abc} \ , \label{LeffIntExp.EQ} \\
        &=& \frac{1}{2\Lambda^2} \left\{
 \left[ \overline{\Dp^c_b} \, \widetilde\Gamma \Dp_a \right] \ \left[ \bar{\Up}^c_c \, \widebar\Gamma^c \chi \right] +
 \left[ \overline{\Dp}_a \, \widebar{\widetilde\Gamma} \Dp^c_b \right] \ \left[  \bar{\chi} \, \Gamma^c \Up^c_c \right]
 \right\} \, \epsilon^{abc} \ , \label{LeffConjIntExp.EQ}
\eea
where
$\widebar{\widetilde\Gamma} \equiv \gamma^0 \widetilde\Gamma^\dagger \gamma^0$,
$\widebar\Gamma \equiv \gamma^0 \Gamma^\dagger \gamma^0$,
$\Gamma^c \equiv C \Gamma^* C$ (where $C=-i\gamma^2$ is the charge-conjugation matrix, see Appendix~\ref{spinorAlg.SEC}),
from which follows $\widebar\Gamma^c = C\gamma^0 \Gamma^T \gamma^0 C$. 
In Appendix~\ref{spinorAlg.SEC},
we show that the $\overline{\Dp^c_a} \widetilde\Gamma \Dp_b  \, \epsilon^{ab...}$
part of $\LInt$ that involves two identical Grassmannian $\Dp$ fields, for the (Majorana-like) $SS$ interaction
is antisymmetric in spin, and imposing the antisymmetry in color also, forces the $SS$ interaction to zero.   
This then means that we cannot write down the $SS$ interaction. 

This leaves only the $VV$ interaction as a possibility for the interaction with two identical $\Dp$ fields,
and writing it out explicitly, we have 
\beq
\LIntVV = \frac{1}{2\Lambda^2} \left[  \overline{\Dp^c_b} \, \gamma^\mu \left(\tilde{g}_L P_L + \tilde{g}_R P_R \right) \Dp_a \right]
    \ \left[ \bar{\chi} \, \gamma_\mu \left( g_L P_L + g_R P_R \right) \Up_c \right]  \, \epsilon^{abc} + h.c. \ . 
\eeq
As shown in Eq.~(\ref{DDspinorRel.EQ}) in Appendix~\ref{spinorAlg.SEC}, the first part of this interaction with two identical $\Dp$ fermions
and with color antisymmetry has a constraint, namely,
$(\overline{\Dp}_a C\,\widetilde\Gamma^* C\, \Dp^c_b) \, \epsilon^{ab...} =
-(\overline{\Dp}_a \gamma^0 \widetilde\Gamma^\dagger \gamma^0 \Dp^c_b) \, \epsilon^{ab...}$.
Applying this to the form in the $VV$ interaction, namely,
$\widetilde\Gamma = \gamma^\mu \left(\tilde{g}_L P_L + \tilde{g}_R P_R \right)$, 
we obtain the constraint $ \tilde{g}_L = \tilde{g}_R \equiv \tilde{g}$.
Taking this into account, we write the $VV$ interaction as
\beq
\LIntVV = \frac{1}{2\Lambda^2} \left[\overline{\Dp^c_b} \, \gamma^\mu \tilde{g} \Dp_a \right]
\ \left[ \bar{\chi} \, \gamma_\mu \left( g_L P_L + g_R P_R \right) \Up_c \right]  \, \epsilon^{abc} + h.c. \ . 
    \label{LeffIntVV.EQ}
\eeq
We thus identify $\widetilde\Gamma = \gamma^\mu \tilde{g} $  and $\Gamma = \gamma_\mu ( g_L P_L + g_R P_R)$
for the $VV$ interaction.
It is this $VV$ interaction that we consider in the remainder of our work.
We note in passing that the Lorentz structure $\overline{\Dp^c} \, \gamma^\mu \Dp$ in the interaction connects different chiralities,
i.e. is of the form $\Dp_L (...) \Dp_R$, written symbolically. 

By a Fierz rearrangement, as explained in Appendix~\ref{Fierz.SEC},
we can equivalently write the VV interaction as a sum over SS operators and other related VV operators (cf. Sec.~\ref{UVcompl.SEC}).
The equivalent Fierz rearranged form we find is
\bea
\LIntVV &=& \frac{1}{2\Lambda^2} \left\{
   \left[ \overline{\Dp^c_b} \, \gamma^\mu g_L P_L\, \Up_c \right]\ \left[ \bar{\chi} \, \gamma_\mu \tilde{g} P_L\, \Dp_a \right]
+  \left[ \overline{\Dp^c_b} \, \gamma^\mu g_R P_R\, \Up_c \right]\ \left[ \bar{\chi} \, \gamma_\mu \tilde{g} P_R\, \Dp_a \right]
\right. \label{VVFierz.EQ} \\ &&\hspace*{1cm} \left.
-2 \left[ \overline{\Dp^c_b} \,           g_L P_L\, \Up_c \right]\ \left[ \bar{\chi} \,            \tilde{g} P_R\, \Dp_a \right]  
-2 \left[ \overline{\Dp^c_b} \,           g_R P_R\, \Up_c \right]\ \left[ \bar{\chi} \,            \tilde{g} P_L\, \Dp_a \right]  
\right\} 
\, \epsilon^{abc}  + h.c. \ , \nonumber
\eea
We see that in addition to the the usual SS operators considered in the literature, 
new related VV operators (cf. Sec.~\ref{UVcompl.SEC}) are also present. 

To our Lagrangian, we add effective mass terms 
\bea
    \LDirMass = - M_{\Dp} \overline{\Dp} \Dp - M_{\Up} \overline{\Up} \Up - M_\chi \overline{\chi} \chi \ , \\
    \LMajMass = - \frac{1}{2} \overline{\chi^c} (\tilde{M}_L P_L + \tilde{M}_R P_R)\, \chi + h.c. \ ,
\label{LMassEff.EQ}    
\eea
where the first line contains Dirac mass terms, while the second line are Majorana mass terms for $\chi$,
with the charge-conjugate field as defined already (see Appendix~\ref{spinorAlg.SEC}).
The details of the origin of these masses are not important for our purposes here.
It is of crucial importance to note that the effective Majorana mass terms $\tilde{M}_{L,R}$, which could be complex,
are the only sources of breaking of baryon number in our effective theory.
In other words, in the limit of zero $\tilde{M}_{L,R}$,
baryon number is a good classical symmetry of the effective theory (including the SM), with the charges as given earlier.
With $\tilde{M}_{L,R}$ nonzero, the Dirac fermion $\chi$ splits up into two Majorana fermions $\Chi_n$, with indefinite baryon number.
Thus, we anticipate that any baryon number violating process must involve the $\Chi_n$ fields and be proportional to the Majorana masses $\tilde{M}_{L,R}$. 

We remove unphysical phases by field redefinitions, and work in the basis where the physical phases are contained in
$g_L \equiv g_{L_0} e^{i\phi_L}$,
$\tilde g_R \equiv \tilde{g}_{R_0} e^{i\tilde\phi_R}$,
$\tilde M_R \equiv \tilde{M}_{R_0} e^{-i\tilde\phi^\prime_R}$,
with $g_{L_0}, \tilde{g}_{R_0}, \tilde{M}_{R_0}$ all real, and all phases $\phi$ also real. 
Furthermore, since $\tilde g_R = \tilde g_L = \tilde g$ for the $VV$ interaction, as shown earlier,
$\tilde g$ has to be real, i.e. $\tilde\phi_R$ is not physical for the $VV$ interaction.
Thus, the phases in $g_L$ and $\tilde M_R$, namely $\phi_L$ and $\tilde\phi^\prime_R$ respectively,
remain as the only two physical phases for the $VV$ interaction, 
and the other parameters, namely,
$g_R$, $\tilde g$, 
$\tilde{M}_L$, $M_\Dp$, $M_\Up$, $M_\chi$ are all real.

In terms of 4-spinors, the $\chi$ mass terms can be assembled into
\beq
\LMass = - \frac{1}{2} \bmat \bar\chi & \overline{\chi^c} \emat \bmat M_\chi & \tilde{M}^*_{R} P_{L} + \tilde{M}^*_{L} P_{R} \\
 \tilde{M}_{L} P_{L} + \tilde{M}_{R} P_{R} & M_\chi \emat
\bmat \chi \\ \chi^c  \emat \ ,
\label{chiMDir.EQ}
\eeq
but recall that $\tilde{M}_L$ is real while $\tilde{M}_R$ can be complex.
This off-diagonal mass matrix should be diagonalized to go to the mass basis.
This diagonalization is achieved by a $2\times 2$ unitary rotation matrix $\Umat$ as explained in Appendix~\ref{AppChiM.SEC}.
As shown there in Eq.~(\ref{chi2ChiURot.EQ}),
we write the Dirac $\chi$ in terms of two mass basis Majorana states $\Chi_n = \{ \Chi_1 \, , \Chi_2 \}$ as
\beq
\chi = (\Umat_{1n} P_L + \Umat^*_{2n} P_R) \Chi_n \ ,
\label{chiMBas.EQ}
\eeq
with mass eigenvalues $M_n = \{M_1, M_2\}$.
The phases in $\Umat$ are all proportional to the one physical Majorana mass phase $\tilde \phi^\prime_R$.

We substitute Eq.~(\ref{chiMBas.EQ}) into Eq.~(\ref{LeffIntVV.EQ}), 
and obtain the $VV$ interaction in the $\chi$ mass basis as
\beq
\LIntVV = \frac{\epsilon^{abc}}{2 \Lambda^2} \left\{
\left[  \overline{\Dp^c_b} \, \widetilde{G}_V^\mu \Dp_a \right] \ \left[ \bar{\Chi}_n \, G^n_{V \mu} \Up_c \right]  +
\left[  \overline{\Dp}_a \, \bar{\widetilde{G}}_V^\mu \Dp^c_b \right] \ \left[ \bar{\Up}_c \, \bar{G}^n_{V \mu} \Chi_n \right]
\right\}  \ ,
\label{LIntVVMB.EQ}
\eeq
where we have
$G^n_{V\mu} \equiv \gamma_\mu \left( g_L\, \Umat^*_{1n} P_L + g_R\, \Umat_{2n}  P_R \right) $,
$\bar{G}^n_{V\mu} = \gamma_\mu \left( g_L^* \, \Umat_{1n} P_L + g_R^* \, \Umat^*_{2n}  P_R \right) $, 
$\widetilde{G}_V^\mu = \bar{\widetilde{G}}_V^\mu = \gamma^\mu \tilde{g}$,
and for future use we define $\ghLn \equiv g_L\, \Umat^*_{1n}$ and $\ghRn \equiv g_R\, \Umat_{2n}$.\footnote{
The connection to Eq.~(\ref{LeffIntExp.EQ}) is given by 
$\widetilde{\Gamma} = \widebar{\widetilde{\Gamma}} \equiv \tilde{g} \gamma^\mu $, 
$\Gamma \equiv G^n_{V\mu}$, 
$\bar\Gamma \equiv \bar{G}^n_{V\mu}$.
}

Under charge conjugation, the three fermionic fields $\psi=\{\Chi,\Dp,\Up\}$ transform as
$\psi \to C \psi^* = \psi^c$, which implies $\psi = C {\psi^c}^*$.    
Substituting this in Eq.~(\ref{LIntVVMB.EQ}), using the relations shown below Eq.~(\ref{LeffConjIntExp.EQ}),
we find an equivalent form written in terms of the charge-conjugated fields that is useful to us, namely,
\beq
\LIntVV = -\frac{\epsilon^{abc}}{2 \Lambda^2} \left\{
\left[  \overline{\Dp^c_b} \, \widetilde{G}_V^\mu \Dp_a \right] \ \left[ \bar{\Up}^c_c \, G^n_{\Lambda \mu} \Chi_n \right]  +
\left[  \overline{\Dp}_a \, \bar{\widetilde{G}}_V^\mu \Dp^c_b \right] \ \left[ \bar{\Chi}_n \, \bar{G}^n_{\Lambda \mu} \Up^c_c \right]
\right\}  \ ,
\label{LIntVVMBconj.EQ}
\eeq
where we have defined $\Gamma^c = \GVn_\mu^c \equiv C \GVn_\mu^* C = -\GZBn_\mu$ and $\bar\Gamma^c = \GVBn_\mu^c \equiv C \GVBn_\mu^* C = -\GZn_\mu$,
with $G^n_{\Lambda \mu} = G^n_{V \mu}|_{(P_L \leftrightarrow P_R)}$ and $\bar{G}^n_{\Lambda \mu} =  \bar{G}^n_{V \mu}|_{(P_L \leftrightarrow P_R)}$
(i.e. $G_\Lambda$ is the corresponding $G_V$ with $P_L$ and $P_R$ interchanged).
We also have  
$\widetilde\Gamma^c = \GtVmuc \equiv C \GtVmuSt C  = -\GtVBmu$ and $\bar{\widetilde\Gamma}^c = \GtVBmuc \equiv C \GtVBmuSt C = -\GtVmu$. 

Under a parity transformation that takes $x=(t,\underbar{x}) \to (t,-\underbar{x}) \equiv \tilde{x}$,
the fermionic fields $\psi=\{\Chi,\Dp,\Up\}$ transform as $\psi(x) \to \eta_a \gamma^0 \psi(\tilde{x}) \equiv \widetilde{\psi}(x)$
(see Appendix~\ref{spinorAlg.SEC}),
and taking $\eta_a = i$ for all three fields, we have $\psi(x) = -i \gamma^0 \widetilde{\psi}(\tilde{x})$.
Substituting this in the charge-conjugated form in Eq.~(\ref{LIntVVMBconj.EQ})
we obtain $\LIntVV$ in terms of the charge and parity transformed fields $\widetilde{\psi}^c(\tilde{x})$ as
\beq
\LIntVV = \frac{\epsilon^{abc}}{2 \Lambda^2} \left\{
\left[  \bar{\widetilde\Dp}_b \, \widetilde{G}_V^\mu \widetilde\Dp^c_a \right] \ \left[ \bar{\widetilde\Chi}_n \, \bar{G}^n_{V \mu} \widetilde\Up^c_c \right]  +
\left[  \bar{\widetilde\Dp}_a \, \bar{\widetilde{G}}_V^\mu \widetilde\Dp^c_b \right] \ \left[ \bar{\widetilde\Up}^c_c \, G^n_{V \mu} \widetilde\Chi_n \right]
\right\}  \ .
\label{LIntVVCPtr.EQ}
\eeq

Comparing Eq.~(\ref{LIntVVMBconj.EQ}) and Eq.~(\ref{LIntVVMB.EQ}) we find that we have $C$-invariance if and only if
$\bar{G}^n_{V \mu} = G^n_{\Lambda \mu}$, i.e. for $\ghLn^* = \ghRn$.
Comparing Eq.~(\ref{LIntVVCPtr.EQ}) and Eq.~(\ref{LIntVVMB.EQ}), we find that we have $CP$ invariance if and only if
$G^n_{V \mu} = \bar{G}^n_{V \mu}$, i.e. for $\ghLn^* = \ghLn$ and $\ghRn^* = \ghRn$.
From this, we reach the following important conclusion:
{\em Since the Sakharov conditions for the generation of the baryon asymmetry requires that both $C$ and $CP$ invariances should be violated,
  the couplings should be such that both these relations are violated,
  i.e. we must have ($\ghLn^* \neq \ghRn$) and ($\ghLn^* \neq \ghLn$ or $\ghRn^* \neq \ghRn$).
  \label{CPviolCond.PG}
}

\begin{figure}
\begin{minipage}{6in}
  \centering
  \raisebox{-0.5\height}{\includegraphics[width=0.25\textwidth]{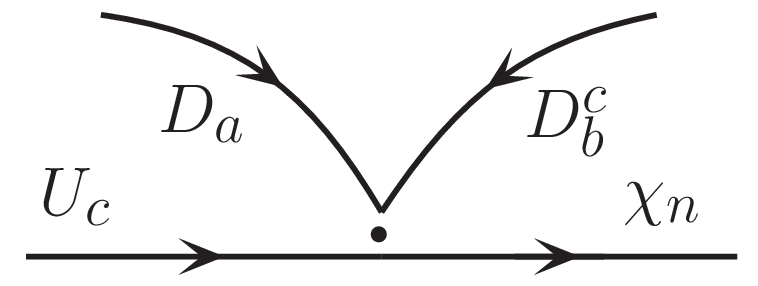}}
  $\phantom{+} \frac{i}{2\Lambda^2} \epsilon^{abc} \widetilde{G}_V^\mu \otimes G^n_{V\mu} $ 
  \hspace*{0.5cm}
  \raisebox{-0.5\height}{\includegraphics[width=0.25\textwidth]{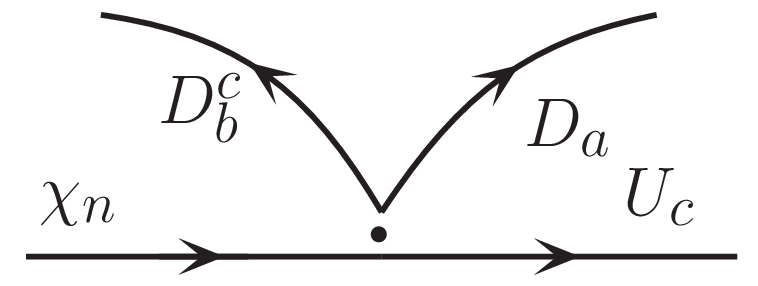}}
  $\phantom{+} \frac{i}{2\Lambda^2} \epsilon^{abc} \bar{\widetilde{G}}_V^\mu \otimes \bar{G}^n_{V\mu} $ 
  \\     \medskip{}
  \raisebox{-0.5\height}{\includegraphics[width=0.25\textwidth]{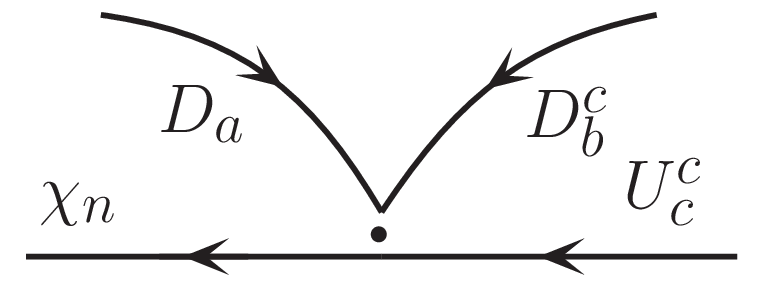}}
  $-\frac{i}{2\Lambda^2} \epsilon^{abc} \widetilde{G}_V^\mu \otimes G^n_{\Lambda\mu} $
  \hspace*{0.5cm}
  \raisebox{-0.5\height}{\includegraphics[width=0.25\textwidth]{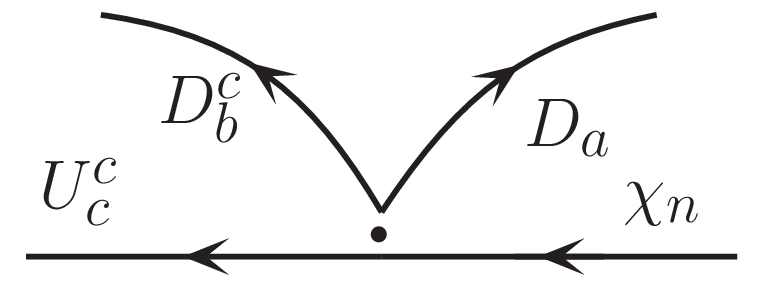}}
  $-\frac{i}{2\Lambda^2} \epsilon^{abc} \bar{\widetilde{G}}_V^\mu \otimes \bar{G}^n_{\Lambda\mu} $
\end{minipage}
\caption{The Feynman rules for the VV interaction, with 
  the first factor showing the coupling for the $\Dp^c\! -\! \Dp$ fermion line, and the second factor for the $\Chi_n \!-\! \Up$ fermion line.
  On the fields, the superscript $c$ denotes the charge conjugate, and the subscript $\{a,b,c\}$ denotes the color index. 
  The direction of the arrow denotes the flow of fermion number (also baryon number).}
\label{FeynRules.FIG}
\end{figure}

\medskip
\noindent\underline{\it Feynman rules:}
From the Lagrangian for the $VV$ interaction in Eq.~(\ref{LIntVVMB.EQ}) and its equivalent form in Eq.~(\ref{LIntVVMBconj.EQ})
we extract the Feynman rules as shown in Fig.~\ref{FeynRules.FIG}. 
We write the couplings as $(...)\otimes (...)$, where the first factor shows the coupling for the $\Dp^c - \Dp$ fermion line,
and the second factor for the $\Chi_n - \Up$ line.
On the fields, the superscript $c$ denotes the charge conjugate, and the subscript $\{a,b,c\}$ denotes the color index,
with the color indices contracted using the $\epsilon^{abc}$ as shown.
The direction of the arrow denotes the flow of fermion number (also baryon number),
and although the Majorana fermions $\Chi_n$ have indefinite baryon number owing to their Majorana mass,
we put an arrow on the $\Chi_n$ leg showing the direction in the limit of zero Majorana mass.

Having the equivalent forms in Eqs.~(\ref{LIntVVMB.EQ}),~(\ref{LIntVVMBconj.EQ}),~and~(\ref{LIntVVCPtr.EQ}) gives us the advantage of
cleanly comparing the process and the charge-conjugated process (or parity-transformed process) with the $u,v$ spinors unchanged between them.
In particular, in computing the decay rate (or scattering rate) difference between the process and (baryon) charge conjugated process to find the baryon asymmetry,
if we use Eq.~(\ref{LIntVVMB.EQ}) to compute the decay rate for the process involving particular $u,v$ spinors,
the decay rate into the opposite baryon charge using Eq.~(\ref{LIntVVMBconj.EQ}) will have the same spinors in the same places,
with only the interaction vertex different.
If we compute the process with any of the $Q\!=\!\{\Up,\Dp\}$ in the final state as
$\bra{Q} (...) = \bra{0}\!a_Q\,(...)$ with ${\cal L}$ of Eq.~(\ref{LIntVVMB.EQ}),
we compute the corresponding conjugate process with $\bar{Q}$ in the final state by picking in the final state
$\bra{\bar{Q}} (...) = \bra{0}\!b_Q\,(...)$ with the ${\cal L}$ of Eq.~(\ref{LIntVVMBconj.EQ}) having the conjugate fields. 
When dealing with conjugate fields (eg. $\Up^c, \Dp^c$) in Feynman diagrams,
we note that as far as the spinors are concerned, the arrow (that shows fermion number) is to be read in the opposite sense for
picking the spinors for those fermion lines,
i.e. in picking the $u,v$ spinors for initial/final state, and for the sign of the mass term in the numerator of their propagators.

\section{UV completions and other effective operators}
\label{UVcompl.SEC}

The non-renormalizable effective interaction of Eq.~(\ref{LeffIntA.EQ}) can be obtained as the lower energy limit of a
renormalizable (UV complete) theory.
Here we consider how this may arise by taking a few example UV completions.
The discussions in this section serve to illustrate in some example UV completions how the effective theory we laid down might arise,
but one should keep in mind that our mechanism of baryon number generation is quite general and not wedded to much of the details in these examples.
Our mechanism of baryon number asymmetry generation assumes that
in addition to the $\chi$ Dirac mass $M_\chi$ that conserves baryon number, 
there is also present Majorana masses $\tilde{M}_{L,R}$ that break baryon number that is responsible for generating a baryon asymmetry.
We leave open the exact origin of this Majorana mass but provide below a simple example for the origin of the Majorana masses.
In addition to demonstrating how the effective operator in Eq.~(\ref{LeffIntA.EQ}) arises in example UV completions, 
we also discuss below how in these UV completions the new physics sector talks to the SM,
which would be relevant for the transfer of the generated baryon asymmetry to the SM.

We assign $\chi$ to have baryon number charge of $B(\chi)=+1$ such that  
baryon number is a good symmetry of our theory in the limit of vanishing $\chi$ Majorana mass.  
In addition to the $\chi$, $\Dp$, and $\Up$ propagating states we have in Sec.~\ref{Th.SEC},
consider, for instance, the addition of an even heavier complex vector field ${\xi}$ in the $\bar{\bf 3}$ of $SU(3)_c$ and having a mass $M_{\xi}$.
We assign baryon number
$B(\xi) = 2/3$,
  with the $\Up,\Dp$ in the fundamental representation of $SU(3)_c$.\footnote{
  Our mechanism of generating the baryon asymmetry goes through for more general EM charge and baryon number assignments,
  and will have the same aspects as long as we have $2 Q(\Dp) + Q(\Up) = 0$ and $2 B(\Dp) + B(\Up) = +1$.
  We take the $\Dp$ and $\Up$ to be in the fundamental of $SU(3)_c$, 
  and we antisymmetrize the interaction in the color index to make them color singlets.
  If some other quantum number antisymmetrizes the interaction,
  our analysis would apply and a baryon number asymmetry can be generated in the same way.}
Next, we discuss in turn two examples UV completions.
  
\medskip
\noindent\underline{\it UV completion A}:
Here, we take $\Dp$ and $\Up$ to be vectorlike quarks, $SU(2)_L$ singlets,
with hypercharge $Y(D) = -1/3$, $Y(U) = 2/3$, and $\chi$ to be uncharged under the SM gauge symmetries. 
We take $\xi$ to be
a singlet of $SU(2)_L$ with $U(1)_Y$ hypercharge $-2/3$ (which implies EM charge $-2/3$),
and denote the state as $\xi^c_{\mu}$ with vector-index $\mu$ and (anti)color index $c$
(the superscript $c$ on this bosonic field should not be confused with the charge-conjugation superscript $c$ on fermionic fields elsewhere).
\begin{table}
  \caption{Quantum number assignments in \textit{UV completion A.}}
\label{UVA.TAB}  
\centering{}%
\begin{tabular}{|c||c|c|c|c|}
\hline 
 & $SU(3)_{C}$ & $SU(2)_{L}$ & $U(1)_{Y}$ & $U(1)_{B}$\tabularnewline
\hline 
\hline 
$\chi$ & 1 & 1 & 0 & 1\tabularnewline
\hline 
$\xi$ & $\bar{3}$ & 1 & -2/3 & 2/3\tabularnewline
\hline 
$U$ & $3$ & 1 & 2/3 & 1/3\tabularnewline
\hline 
$D$ & $3$ & 1 & -1/3 & 1/3\tabularnewline
\hline 
\end{tabular}
\end{table}
We show in Table~\ref{UVA.TAB} these quantum number assignments, 
and consistent with this, we can write down the interactions
\beq
    {\cal L}^{(A)}_{UV} \supset - \frac{1}{2}\, \epsilon^{abc} \overline{\Dp^c_b} \, \tilde{g} \gamma^\mu \Dp_a\, {\xi^c_\mu}^*
    - \bar{\chi} \, \gamma^\mu (g_L P_L + g_R P_R) \Up_c \, \xi^c_\mu + {\rm h.c.} \ .
\label{LUVA.EQ}    
\eeq
We take the BSM scale, i.e. the $\Up,\Dp,\chi$ mass scales, to be TeV scale or higher, leaving the SM structure intact
with the BSM sector decoupling as the BSM scale is raised.
Furthermore, we include a mass mixing between the BSM $\Dp,\Up$ with the SM quarks,
$SU(2)_L$ singlets $u^i_R,~d^i_R$, or doublets $q^i_L$,
where $i$ is the generation index, i.e. $d^i = (d,s,b)$ and $u^i = (u,c,t)$.
Our phenomenological analysis primarily focuses on this scenario.
Before we delve into this further, we briefly comment on the consistency of not including operators 
like those in Eq.~(\ref{LUVA.EQ}) but with SM chiral quarks $q_L,u_R,d_R$, instead of the vectorlike $Q=\{\Up,\Dp\}$.

For an $SU(2)$ singlet $\xi$, the analog of the first term cannot be written down with SM chiral fields
since it would involve both $(d_L,d_R)$ and $SU(2)_L$ invariance would forbid it.
However, the analog of the second term of the form $(\bar{\chi} u_R \xi)$ {\em can} be.
Even in this case,
the effective operator generated upon integrating out $\xi$ would still be suppressed by the $\Lambda \equiv M_\xi$ scale, 
and the baryon asymmetry would still be generated at the $M_\chi$ scale.
(We note that our baryon asymmetry generation mechanism must involve the first operator also since the effective operator gotten via only
the second operator is self-conjugate and will not contain a $CP$-violating phase.)
The presence of the $(\bar{\chi} u_R \xi)$ operator may in fact present an opportunity to probe this in experiments
(such as the \nnbar oscillations discussed in Sec.~\ref{DelB2.SEC})
since it now involves the SM $u_R$.
The flip-side of this is that constraints on the theory would be tighter.
To be safe from constraints, we could forbid the $(\bar{\chi} u_R \xi)$ operator but allow the $(\bar{\chi} \Up \xi)$ operator
by invoking a $Z_3$ symmetry under which the BSM fields transform but the SM fields are singlets.
Under the $Z_3$ let $Q\to e^{i(2\pi/3)\alpha_Q} Q$, $\chi\to e^{i(2\pi/3)\alpha_\chi} \chi$, $\xi\to e^{i(2\pi/3)\alpha_\xi} \xi$,
where the $\alpha_i$ are integer charges. 
$Z_3$ invariance of ${\cal L}^{(A)}_{UV}$ follows if $\alpha_\xi = 2 \alpha_Q$ and $\alpha_\chi = 3\alpha_Q$.
One possible choice is $\alpha_Q = 1$, which yields $Q\to e^{i(2\pi/3)} Q$, $\chi\to \chi$, $\xi\to e^{i2(2\pi/3)} \xi$.
Clearly, this allows ${\cal L}^{(A)}_{UV}$ but forbids the $(\bar{\chi} u_R \xi)$ operator as we intend.
For this assignment, interestingly, the $\chi$ is a $Z_3$ singlet, and the Majorana mass generation mechanism (discussed below)
is unaffected by this consideration.
Our identification of the $Z_3$ symmetry implies that leaving out the corresponding operators with the SM quarks is consistent
in the sense of Ref.~\cite{tHooft:1979rat}, i.e. radiatively stable. 

\begin{figure}
\begin{minipage}{6in}
  \centering
  \raisebox{-0.5\height}{\includegraphics[width=0.25\textwidth]{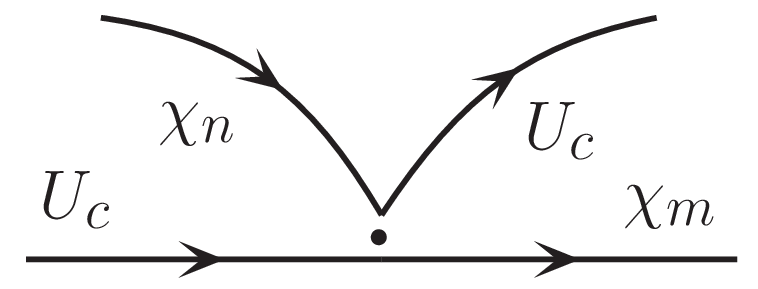}}
  $-\frac{i}{\Lambda^2} \bar{G}_V^{n\mu} \otimes G^m_{V\mu} $ 
  \hspace*{1cm}
  \raisebox{-0.5\height}{\includegraphics[width=0.25\textwidth]{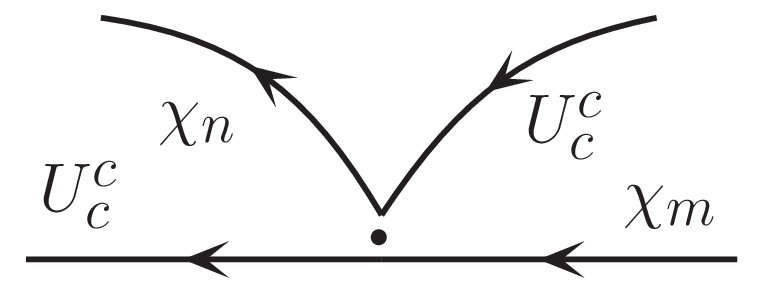}}
  $-\frac{i}{\Lambda^2} G_\Lambda^{n\mu} \otimes \bar{G}^m_{\Lambda\mu} $ 
  \\     \medskip{}
  \hspace*{3cm}
  \raisebox{-0.5\height}{\includegraphics[width=0.25\textwidth]{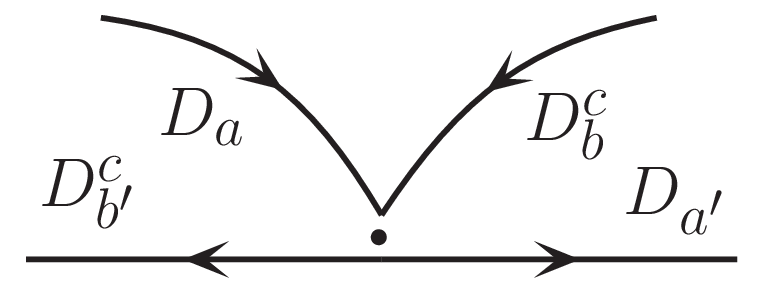}}
  $\frac{i}{2\Lambda^2} (\delta^{aa'} \delta^{bb'} - \delta^{ab'} \delta^{ba'}) \widetilde{G}_V^\mu \otimes \bar{\widetilde{G}}_{V\mu}$
\end{minipage}
\caption{The Feynman rules of the other operators associated with the VV interaction,
  with each factor associated with the corresponding fermion line. 
  On the fields, the superscript $c$ denotes the charge conjugate, and the subscript $\{a,b,c\}$ denotes the color index. 
  The direction of the arrow denotes the flow of fermion number (also baryon number).}
\label{FeynRulesModA.FIG}
\end{figure}

For physics below the $M_{\xi}$ scale, we can use an effective theory obtained by integrating out the $\xi$ field,
which leads to new dimension-six effective operators.
Doing so, we generate the effective operator of Eq.~(\ref{LeffIntVV.EQ}) with $\Lambda \equiv M_{\xi}$,
which in the mass basis is Eq.~(\ref{LIntVVMB.EQ}).
In addition, we also generate the effective operators
\beq
    {\cal L} \supset - \frac{1}{\Lambda^2} \, [\bar{\Up}_c \bar{G}^n_{V \mu} \Chi_n]\, [\overline{\Chi}_m {G^m_{V}}^\mu \Up_c] 
    - \frac{1}{2 \Lambda^2} (\delta^{aa'} \delta^{bb'}
    - \delta^{ab'} \delta^{ba'}) [\overline{\Dp}^c_b\, \tilde{g} \gamma^\mu \Dp_a]\, [\bar{\Dp}_{a'}\, \tilde{g} \gamma_\mu \Dp^c_{b'}] \ ,
\label{modAops.EQ}    
\eeq
which are self-conjugate.
Such operators have been considered in the literature, albeit with the scalar-scalar Lorentz structure,
while our operators above are of the vector-vector type.
Writing the first term in Eq.~(\ref{modAops.EQ}) in terms of the conjugate fields, we can write an equivalent form
$
{\cal L} \supset - (1/\Lambda^2) \, [\bar{\Up}^c_c G^n_{\Lambda \mu} \Chi_n]\, [\overline{\Chi}_m {\bar{G}^{m\mu}_{\Lambda}} \Up_c^c] \ .
$
From these, we can extract the Feynman rules of these new operators as shown in Fig.~\ref{FeynRulesModA.FIG}. 
Although one could take the effective couplings of the operators in Eqs.~(\ref{LIntVVMB.EQ})~and~(\ref{modAops.EQ}) as unrelated,
the effects generated by doing so could be misleading as a particular UV completion could indeed relate them in a definite way as we see here. 

Other operators that could be generated in a different UV completion,
for instance by integrating out a different heavier state of EM charge $1/3$,
are
$[\overline{\Dp^c} \Gamma \Up_c]\, [\bar{\chi} \Gamma \Dp]$,   
$[\overline{\Dp^c} \Gamma \Up_c]\, [\bar{\Up} \Gamma \Dp^c]$,  
and
$[\overline{\Dp} \Gamma \chi]\, [\bar{\chi} \Gamma \Dp]$,     
where we show the Lorentz structure symbolically as $\Gamma$. 
Some of these will be contained in a Fierz rearrangement of the earlier operators. 
Since the origin of these operators are different, it would not be unreasonable to consider these operators as 
uncorrelated with the earlier couplings, and to keep the analysis manageable,
we omit these and only consider the operators in Eqs.~(\ref{LIntVVMB.EQ})~and~(\ref{modAops.EQ}) in our study.
Furthermore, for reasons that we explain in a follow-up work~\cite{OurBGChiFeynNum.BIB},
we find
that for our choices of masses,
the baryon asymmetry due to including the operators in Eq.~(\ref{modAops.EQ}) is subdominant and we therefore mainly focus on the operator
in Eq.~(\ref{LIntVVMB.EQ}) in our work.

Next, we discuss one way in which the new physics sector could couple to the SM.  
The $\Dp$ and/or the $\Up$ BSM fields, having SM-like EM charges,
could have mass mixing terms with any of the SM quark flavors of the same EM charge.
  The $\Dp$ and $\Up$ can interact with the SM via the interaction terms
  \beq
      {\cal L}_{\rm Yuk} \supset -\tilde{y}_d \bar\Dp H^\dagger q_L + \tilde{y}_u \bar\Up H \cdot q_L + {\rm h.c.} \ ,
      \label{LuUdDmix.EQ}
  \eeq
  where $q_L$ is the SM $SU(2)_L$ doublet quark of any generation (we suppress generation index),
  $H$ is the SM Higgs doublet,
  and $H\cdot q_L$ represents the antisymmetric combination of the $SU(2)_L$ indices.\footnote{Proton stability
  motivates us to keep lepton number symmetry unbroken at the classical level, 
  and therefore we do not include the $\bar\chi L\cdot H$ operator that breaks lepton number, where the $L$ is an $SU(2)$ leptonlike doublet,
  and we assign lepton number 0 to $\chi$.}
  After electroweak symmetry breaking (EWSB) this interaction induces mass mixing terms $\bar\Dp_R d_L$ and $\bar\Up_R u_L$, 
  which are diagonalized to go to the mass basis with
  the sine of the mixing angles $s_{d\Dp}\equiv\sin{\theta_{d\Dp}}$, $s_{u\Up}=\sin{\theta_{u\Up}}$
  and mass eigenstates $d_{1,2}$ and $u_{1,2}$ as detailed in Refs.~\cite{Gopalakrishna:2011ef,Gopalakrishna:2013hua}.
  For notational ease, we continue denoting the lighter mass eigenstate as $d,u$ and the heavier mass eigenstate as $\Dp,\Up$. 
  This is abuse of notation, but in the limit of small mixing, and to leading order in the mixing, this is appropriate as we have $u_1 \approx u$ and $u_2 \approx \Up$,
  and similarly for the $d,\Dp$.
  We note that the $\tilde{y}_{u,d}$ breaks the $Z_3$ symmetry identified above. 
  In the limit $\tilde{y}_{u,d} \to 0$, this $Z_3$ becomes exact, and therefore taking
  $\tilde{y}_{u,d}$ small is radiatively stable, i.e. this is technically natural~\cite{tHooft:1979rat}. 
  The bottom-line is that the mass-mixing terms induces the decays $\Dp \to (d h, d Z, u W^-)$ and $\Up \to (u h, u Z, d W^+)$,
  where $h$ is the Higgs boson, and $W^\pm, Z$ are the SM electroweak gauge bosons.
  These decays are discussed in detail in Refs.~\cite{Gopalakrishna:2011ef,Gopalakrishna:2013hua} and references therein.  
  These decays being $B$ conserving, the baryon asymmetry contained in $\Dp,\Up$ is transferred fully to the SM $d,u$ in these decays.
  This provides a concrete example in which the baryon asymmetry generated and contained in the $\Dp,\Up$ sector could be transferred into the SM quarks.

\medskip
\noindent\underline{\it UV completion B}:
Here we consider the possibility that one or both of the $\Dp$, $\Up$ fields could be taken to be any of the
$d^i = (d,s,b)$ and/or $u^i = (u,c,t)$ SM fields respectively, with $i$ being the generation index.
As noted below Eq.~(\ref{LeffIntVV.EQ}),
the $\overline{\Dp^c} \, \gamma^\mu \Dp$ part of the effective VV interaction in Eq.~(\ref{LeffIntA.EQ})
connects different Lorentz chiralities since it is of the form $\widebar{\Dp^c}_{\!L,R}\, \gamma^\mu (...) \Dp_{R,L}$.
This implies that if the $\Dp$ is taken to be the SM $d$ quark, one of these must be a $d_L$ and contained in $q_L$ for $SU(2)_L$ invariance, 
and must necessarily also involve the $u_L$.
  We ensure the $SU(2)_L$ invariance of the UV completion by taking the $\xi_\mu^c$ to be a doublet. 
  This allows two alternatives, which we refer to as {\it UV completions} {\it B$_1$} and {\it B$_2$},
  depending on whether the $\chi$ is a singlet or doublet of $SU(2)_L$ respectively.
We show in Table~\ref{UVB.TAB} the quantum number assignments for {\it UV completion $B_1$}~(left) and {\it $B_2$}~(right).  
\begin{table}
\caption{Quantum number assignments in \textit{UV completion $B_{1}$}(left)
  and \textit{UV completion $B_{2}$}(right).}
\label{UVB.TAB}
\centering{}%
\begin{tabular}{|c||c|c|c|c|}
\hline 
 & $SU(3)_{C}$ & $SU(2)_{L}$ & $U(1)_{Y}$ & $U(1)_{B}$\tabularnewline
\hline 
\hline 
$\chi$ & $1$ & $1$ & $0$ & $1$\tabularnewline
\hline 
$\xi$ & $\bar{3}$ & $2$ & $-1/6$ & $2/3$\tabularnewline
\hline 
$Q$ & $3$ & $2$ & $1/6$ & $1/3$\tabularnewline
\hline
$D$ & $3$ & $1$ & $-1/3$ & $1/3$\tabularnewline
\hline 
\end{tabular}~~~~~~%
\begin{tabular}{|c||c|c|c|c|}
\hline 
 & $SU(3)_{C}$ & $SU(2)_{L}$ & $U(1)_{Y}$ & $U(1)_{B}$\tabularnewline
\hline 
\hline 
$\chi$ & $1$ & $2$ & $-1/2$ & $1$\tabularnewline
\hline 
$\xi$ & $\bar{3}$ & $2$ & $-1/6$ & $2/3$\tabularnewline
\hline 
$Q$ & $3$ & $2$ & $1/6$ & $1/3$\tabularnewline
\hline 
$U$ & $3$ & $1$ & $2/3$ & $1/3$\tabularnewline
\hline 
$D$ & $3$ & $1$ & $-1/3$ & $1/3$\tabularnewline
\hline 
\end{tabular}
\end{table}
We discuss each one in turn next. 

  First, in {\it UV completion B$_1$},
  we take $\xi$ to be a doublet of $SU(2)_L$, $\chi$ to be a singlet, include an $SU(2)_L$ doublet $Q$ and a singlet $D$,
  and write
\beq
    {\cal L}^{(B_1)}_{UV} \supset - \frac{1}{2}\, \epsilon^{abc}\, {\xi_\mu^c}^\dagger \overline{Q^c_b} \, \gamma^\mu (\tilde{g}_L P_L + \tilde{g}_R P_R) \Dp_a 
                               - \bar{\chi} \, \gamma^\mu (g_L P_L + g_R P_R)\, Q_c\!\cdot\!\xi_\mu^c + {\rm h.c.} \ ,
\eeq
where the $(\cdot)$ denotes the antisymmetric combination of the $SU(2)_L$ indices.                               
We assign the
hypercharges $Y(Q) = 1/6$, $Y(\chi) = 0$, $Y(\xi) = -1/6$,
and baryon number $B(Q) = 1/3$. 
We can write the doublet components as $Q = (Q_u \ Q_d)^T$ with $Q_u$ having EM charge +2/3 and $Q_d$ having EM charge -1/3,
and $\xi = (\xi_{\frac{1}{3}} \ \xi_{-\frac{2}{3}})^T$ showing the EM charge of the components as subscripts.
  
Next, in {\it UV completion B$_2$},
we take $\xi$ to be a doublet of $SU(2)_L$, $\chi$ to be a doublet, include a doublet $Q$,
  and with the $\Up,\Dp$ as before, 
  we write
\beq
    {\cal L}^{(B_2)}_{UV} \supset - \frac{1}{2}\, \epsilon^{abc}\, {\xi_\mu^c}^\dagger\, \overline{Q^c_b} \, \gamma^\mu (\tilde{g}_L P_L + \tilde{g}_R P_R) \Dp_a 
                               - \bar{\chi} \, \gamma^\mu (g_L P_L + g_R P_R) \Up_c \, \xi^c_\mu + {\rm h.c.} \ .
\eeq
We assign the
hypercharges $Y(Q) = 1/6$, $Y(\chi) = -1/2$, $Y(\xi) = -1/6$,
and baryon number $B(Q) = 1/3$. 
We can write the doublet components as $Q=(Q_u \ Q_d)^T$, $\chi = (\chi_+ \ \chi_0)^T$ and $\xi = (\xi_{\frac{1}{3}} \ \xi_{-\frac{2}{3}})^T$,
showing the EM charge of the $\chi, \xi$ components as subscripts.
The mechanism of $U(1)_B$ breaking in the UV completion must ensure that only $\chi_0$ receives a Majorana mass in order to not spoil EM invariance. 

We make the specific choices in developing UV completion {\it B$_2$} as it allows us the
possibility to take the $\Dp$, $\Up$, and $Q$ to be the chiral SM quarks, namely,
we can identify $\Dp \to d_R^i$, $\Up \to u_R^i$, and $Q \to q_L^i$.
This now allows chiral couplings $\tilde{g}_{L,R}$, unlike in UV completion {\it A} which has vectorlike couplings $\tilde{g}$.
This gives us a realization in which it may be possible to generate the baryon asymmetry directly in the SM sector,
while in UV completion {\it A}, the baryon asymmetry is generated in the BSM sector and transferred to the SM by mass mixing, as already discussed. 

Integrating out the $\xi$ fields in the UV completions {\it B$_1$} and {\it B$_2$} leads to lower energy effective theories containing
the effective operator in Eq.~(\ref{LIntVVMB.EQ}) along with other associated operators.
The details of which other operators are generated in UV completion {\it B} is left for future work.
Since the $\Up,\Dp$ are embedded in the SM sector, there is no clean separation of scales between the SM and BSM sectors,
and the phenomenological implications of coupling the $\chi$ to the SM sector will have to be worked out more carefully.  
We do not do this here, but only present this UV completion possibility as something that could be developed further. 
In the following, in working out the baryon asymmetry generated, we primarily have in mind UV completion {\it A}
and take the $\Up,\Dp$ to be vectorlike quarks with masses bigger than a TeV;
however, we expect that our mechanism of baryon asymmetry generation should work
even if the $\Up,\Dp$ are identified with the SM chiral quarks. 

\medskip
\noindent\underline{\it Majorana mass generation}:
Here, we illustrate with a simple example how the baryon number violating $\chi$ Majorana mass might arise
from a baryon number conserving theory.
For this, we introduce a complex scalar field $\Phi_B$ with Yukawa couplings to the $\chi$,
and write down a baryon number conserving Lagrangian density
\beq
{\cal L} = - M_\chi \bar\chi \chi - \frac{1}{\sqrt{2}}\, \left[\Phi_B\, \widebar{\chi^c} (\tilde{y}_L P_L + \tilde{y}_R P_R) \chi + h.c.\right] - {\cal V}(\Phi_B) \ ,
\eeq
where $M_\chi$ is the Dirac mass.
Recalling the baryon number assignment $B(\chi) = +1$, we assign $B(\Phi_B) = -2$ for invariance under $U(1)_B$. 
If the potential energy density for the $\Phi_B$ field, ${\cal V}(\Phi_B)$,
is such that the minimum of the potential is at a nonzero value $\left<\Phi_B\right> = v_B/\sqrt{2} \neq 0$,
baryon number is spontaneously broken, resulting in Majorana mass terms $\tilde{M}_{L,R} = y_{L,R} v_B$,
the effective Majorana masses written down in Eq.~(\ref{LMassEff.EQ}).
The spontaneous breaking of a global symmetry, i.e. $U(1)_B$ here,
implies the presence of a massless Nambu-Goldstone boson (NGB) which is problematic phenomenologically.
A discussion of the mechanism that gives a mass to the NGB alleviating this problem is beyond the scope of our work, 
and this naive example only serves to illustrate how the baryon number violating Majorana masses
might arise from a $U(1)_B$ conserving theory. 

  We turn next to discussing our mechanism for generating the baryon asymmetry at the $\Dp$, $\Up$ level,
  being agnostic to the details of how the so generated asymmetry is transferred to the SM sector.

\section{Generating the Baryon Asymmetry}
\label{ABgen.SEC}

In this section, we discuss the mechanism of generation of a baryon asymmetry in our effective theory
we laid out in Sec.~\ref{Th.SEC} that couples the $\Chi_n$ with the $Q=\{\Up,\Dp\}$.
First, we consider the baryon asymmetry from the decay of the Majorana $\Chi_n$  
due to the decay rate difference between $\Gamma(\Chi_n \to QQQ)$ and $\Gamma(\Chi_n \to \bar{Q}\bar{Q}\bar{Q})$,
where the former decay final state has baryon number $B=+1$, while the latter has $B=-1$. 
Second, we consider the baryon asymmetry from $\Chi_n$ scattering on a $\bar{Q}$ or $Q$
due to a difference between the scattering cross section $\sigma(\Chi_n \bar{Q} \to QQ)$ and $\sigma(\Chi_n Q \to \bar{Q}\bar{Q})$,
where the former process has $\Delta B = +1$ while the latter has $\Delta B = -1$.
We discuss each of these in turn, next. 
In our discussion, we only consider unpolarized rates,
i.e. take it that the matrix element squared are averaged over initial-state spins and summed over final-state spins.  

\subsection{Baryon Asymmetry from Decay}
\label{ABdec.SEC}

The Majorana fermions $\Chi_n$ ($n=\{1,2\}$) have indefinite baryon number and can decay either to
$\Dp \Dp \Up$ or to $\bar\Dp \bar\Dp \bar\Up$ final states which have $B=+1$ and $-1$, respectively.
Equivalently, in our theory, the baryon asymmetry arises due to a difference in the decay rates
$\Gamma^n \equiv \Gamma(\Chi_n \to \widebar{\Dp^c} \Dp \Up)$
vs. 
$\Gamma^{c n} \equiv \Gamma(\Chi_n \to \widebar{\Dp} \Dp^c \Up^c)$,
where the former decay has a final state with $B=+1$ while the latter $B=-1$.  
Since $\Chi_n$ is a Majorana particle, i.e. $\Chi_n^c = \Chi_n$, the initial state is the same in both, but decaying to final states
with opposite $B$.

We denote, for each $n$, the amplitude for $\Chi_n \to \widebar{\Dp^c} \Dp \Up$ as
$\ampA^n = \ampA_0^n + \ampA_1^n + ...$ with $\ampA_0^n$ being the tree-level amplitude and $\ampA_1^n$ the loop-level amplitude,
and for the conjugate process $\Chi_n \to \widebar{\Dp} \Dp^c \Up^c$ as $\ampA^{cn} = \ampA^{cn}_0 + \ampA^{cn}_1 + ...$~.
The decay rate for the process is then given by
\beq
\Gamma^n = \frac{1}{2 M_n} \int d\Pi_3\, |\ampA^n|^2 \ ,
\label{Gm0.EQ}
\eeq
where we integrate over the three-body final-state phase-space $d\Pi_3$,
and for the conjugate process decay, the $\Gamma^{c n}$ is given similarly in terms of $|\ampA^{c n}|^2$.
We define the baryon asymmetry $\AsymB$ as 
\beq
\AsymB = \sum_{n=1,2} \AsymB^n \ , \ {\rm with}\ \, 
\AsymB^n \equiv \frac{\Gamma^n - \Gamma^{c n}}{\Gamma^n + \Gamma^{c n}} \ , \quad
 \ .
\label{AsymBDefn.EQ}
\eeq
Not surprisingly, we find that at tree level, i.e. leading order (LO), in the $VV$ interaction, 
the decay rate is the same for the process and its conjugate process, i.e. $\Gamma_0 = \Gamma_0^c$.
So we must go to loop level, i.e. at least to the next to leading order (NLO) level,
to possibly have a nonzero $\AsymB$.
In the following, to avoid clutter, we suppress the $n$ on these amplitudes, but it is understood where appropriate.

Now, $\Gamma \, \propto\, |\ampA|^2$ and $\Gamma^c \, \propto\, |\ampA^c|^2$
from which it is easy to see that
$\AsymB\, \propto\, [ (\ampA_0^* \ampA_1 - \ampA_0^{c*} \ampA_1^c) + h.c. ] + ... $
where we keep only the terms to lowest order in the $VV$ coupling that give a nonzero asymmetry.
Thus, the interference term between tree and loop contributions is the leading contribution to $\AsymB$, and we have for each $n$
\beq
\AsymB\, = \, \frac{2\, \Delta \Gamma_{01}}{(\Gamma + \Gamma^c)}
           \approx \frac{\Delta \Gamma_{01}}{\Gamma_0} \ , \label{AsymBA0A1.EQ} 
            {\rm where}\
           \Gamma_0 = \frac{1}{2 M_n} \int d\Pi_3 \ |\ampA_0|^2 \ ; \quad
           \Delta \Gamma_{01} =  \frac{1}{2 M_n} \int d\Pi_3 \ {\rm Re}(\ampA_{01} - \ampA_{01}^c)  \ ,
\eeq
with $\ampA_{01}\equiv \ampA_1 \ampA_0^*$ and $\ampA_{01}^c \equiv \ampA_1^c \ampA_0^{c*}$,
we include $\int d\Pi_3$ for the integration over the 3-body phase space,
and,
$\Delta\Gamma_{01} = \Gamma_{01} - \Gamma_{01}^c$ where 
$\Gamma_{01} = (1/(2M_n)) \int d\Pi_3 \, {\rm Re}(\ampA_{01})$ and $\Gamma_{01}^c$ is with $\ampA_{01}^c$.
Since there is no asymmetry at tree level, in the denominator we have taken $\Gamma \approx \Gamma^c \approx \Gamma_0$
for obtaining the leading contribution to $\AsymB$. 

The interaction vertices $G^n_{V\mu}$ in Eq.~(\ref{LIntVVMB.EQ}) that enter in this process
are complex due to the phases in the elements of $\Umat$ (due to a phase in $\tilde M_R$, i.e. $\tilde\phi^\prime_R$), 
and in the coupling $g_L$ (phase being $\phi_L$).
In general terms, we can write the tree and loop amplitudes
for the process as $\ampA_{0,1} \equiv a_{0,1} e^{i \phi_{0,1}}$,
and for the conjugate process as $\ampA^c_{0,1} \equiv a^c_{0,1} e^{i \phi^c_{0,1}}$,
where the $\phi_{0,1}$ are combinations of phases coming from the couplings.
The interference term for the process is $\ampA_{01} = \ampA_1 \ampA_0^* = a_{01} e^{i\phi_{01}}$,
where $a_{01} = a_0 a_1$ and $\phi_{01} = \phi_1 - \phi_0$,
and similarly for the conjugate process we have 
$\ampA^c_{01} = \ampA^c_1 {\ampA_0^c}^* = a^c_{01} e^{i\phi^c_{01}}$
with 
$a^c_{01} = a^c_0 a^c_1$ and $\phi^c_{01} = \phi^c_1 - \phi^c_0$.
Assuming for now that no intermediate particles go on shell, the only phases are those in the couplings. 
From the way the couplings enter the amplitudes,
we find that the conjugate process amplitudes $\ampAc_0, \ampAc_1$ (and $\ampA_{01}^c$) have
the same corresponding magnitudes but opposite phases as the corresponding process,
i.e. $a_{0,1}^c = a_{0,1}$ and $\phi_{0,1}^c = - \phi_{0,1}$,
and the interference term for the conjugate process evaluates to $a_{01} e^{-i\phi_{01}}$.
The $\phi_0,\phi_1$ (and $\phi_{01}$), that flip sign in going from the process to the conjugate process,
are commonly referred to as ``weak-phases''.
Equation~(\ref{AsymBA0A1.EQ}) then tells us that
$\AsymB\, \propto\, 2a_{01} \,{\rm Re}((e^{i\phi_{01}} - e^{-i\phi_{01}}) = 0$; in other words we find no asymmetry generated here.

However,
if intermediate states can go on shell and the loop diagram can be cut, 
a discontinuity arises in the amplitude,
and a piece $i \hat\ampA_1 \equiv (1/2)~{\rm Disc}(\ampA_1)$ is added to the amplitude~\cite{Peskin:1995ev}.
The $\hat\ampA_1$ itself may be complex due to phases in the couplings.
The above factor of $i$, i.e. an extra phase of $\pi/2$,
comes with the {\em same sign} for the process and the conjugate process,
and is like the ``strong phase'' (usually denoted as $\delta$).\footnote{
  In fact, given that the final state is colored leading to hadrons in the final state,
  their rescattering via (QCD) strong interactions could also lead to a phase that is the same for the process and its conjugate process.
  We do not include such a strong phase $\delta$ here, but if we do, all our expressions would include an extra factor of $\sin\delta$.
Our expressions are written down for the special case of $\delta = \pi/2$.}
Correspondingly, the discontinuity adds to the interference term $\ampA_{01}$
the piece $i \hat\ampA_{01} = i\hat\ampA_1 \ampA_0^*$,
and we write its contribution to the decay rate as $i\hat\Gamma_{01}$.
Similarly, for the conjugate process, the discontinuity adds in the interference term the piece $i\hat\Gamma_{01}^c$.

In general terms, in the loop amplitude $\ampA_1$, we can write the piece added due to the cut as
$i\hat\ampA_1 = a_1 e^{i (\phi_1+\pi/2)}$
and for the conjugate process as
$i\hat\ampA^c_1 = a^c_1 e^{i (\phi^c_1+\pi/2)}$. 
Thus, for the process, the piece added due to the cut to the interference term $\ampA_{01}$ is
$i\hat\ampA_{01} = a_0 a_1 e^{i(\phi_{01}+\pi/2)}$,
and for the conjugate process it is
$i\hat\ampA^c_{01} = a^c_0 a^c_1 e^{i(\phi^c_{01}+\pi/2)}$.
If there is no asymmetry at tree level, we have $a_0 = a_0^c$, in which case, from Eq.~(\ref{AsymBA0A1.EQ}),
we obtain 
\beq
\AsymB^n \approx - \frac{\Delta\hat\Gamma^n_{01}}{\Gamma^n_0} \ , \quad
       {\rm where}\ \Delta\hat\Gamma_{01} = \frac{1}{2 M_n} \int d\Pi_3 \ {\rm Im}(\hat\ampA_{01} - \hat\ampA^c_{01}) 
   = \frac{1}{2 M_n} \int d\Pi_3 \ a_0 (a_1 \sin{\phi_{01}} - a^c_1 \sin{\phi^c_{01}) }  \ ,
\label{AsymBhat.EQ}
\eeq
and, in addition to $a_0 = a^c_0$ if we also have $\phi^c_0 = -\phi_0$, $a_1 = a^c_1$, and, $\phi^c_1 = -\phi_1$, we obtain
\beq
\Delta\hat\Gamma_{01} 
= \frac{1}{2 M_n} \int d\Pi_3 \ 2\, {\rm Im}(\hat\ampA_{01})
= \frac{1}{2 M_n} \int d\Pi_3 \ (2 a_0 a_1 \sin{\phi_{01}})  \ ,
\label{Gmhat01.EQ}
\eeq
with
$\Delta\hat\Gamma_{01} = \hat\Gamma_{01} - \hat\Gamma_{01}^c$ where 
$\hat\Gamma_{01} = (1/(2M_n)) \int d\Pi_3 \, {\rm Im}(\hat\ampA_{01})$ and $\hat\Gamma_{01}^c$ is with $\hat\ampA_{01}^c$.

This then is a mechanism by which a nonzero baryon asymmetry could be generated in this theory.
This is an instance of well known aspects (see for example Refs.~\cite{Branco:1999fs,Valencia:1994zi}) 
of how an asymmetry arises from the interference of two complex amplitudes ($\ampA_0$ and $\ampA_1$),
with a nonzero weak phase difference ($\phi_{01}$ from couplings) and a nonzero strong phase ($\delta = \pi/2$ from the cuts).

\begin{figure}
\begin{minipage}{6in}
  \centering
  \[
  \AsymB^n \hspace*{0.1cm} \propto \hspace*{0.1cm} 
  \left|
  \begin{array}{c}
    \raisebox{-0.18\height}{\includegraphics[width=0.175\textwidth]{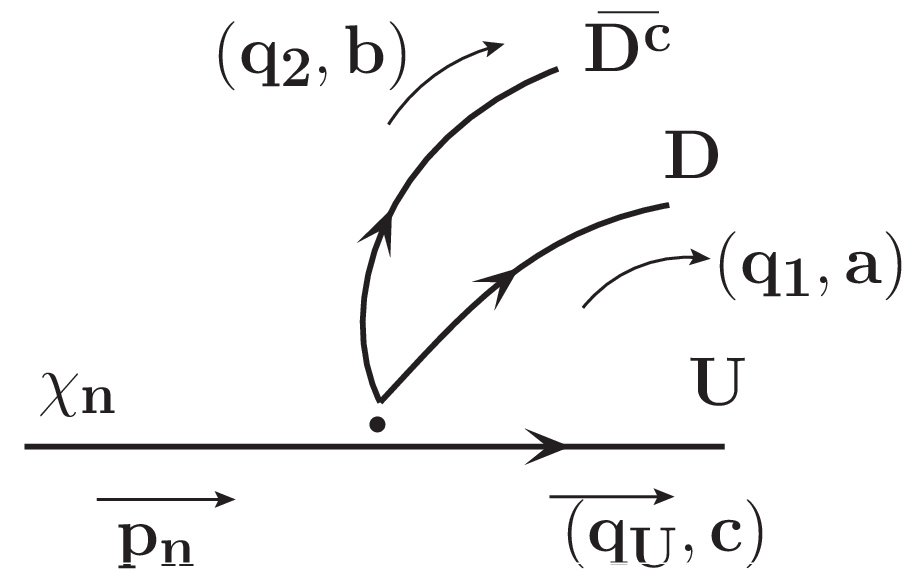}}
    +
    \raisebox{-0.25\height}{\includegraphics[width=0.225\textwidth]{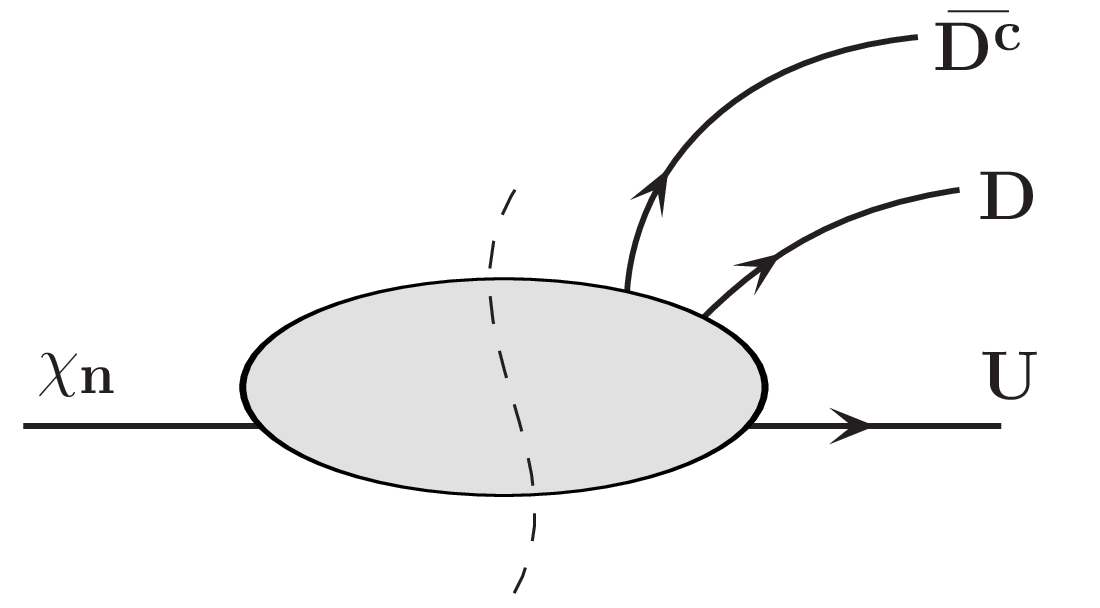}}
  \end{array}  
  \hspace*{-0.25cm} \right|^2
  -
  \hspace*{0.1cm} \left|
  \begin{array}{c}
    \raisebox{-0.18\height}{\includegraphics[width=0.175\textwidth]{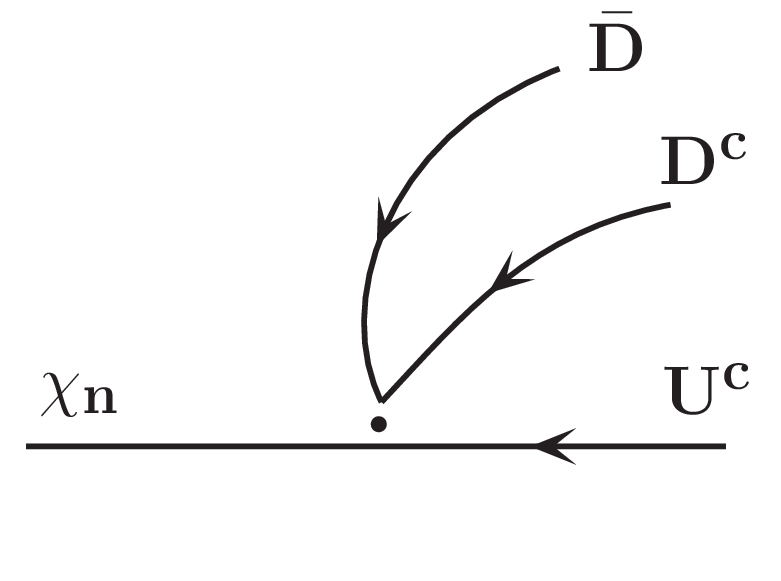}}
    +
    \raisebox{-0.25\height}{\includegraphics[width=0.225\textwidth]{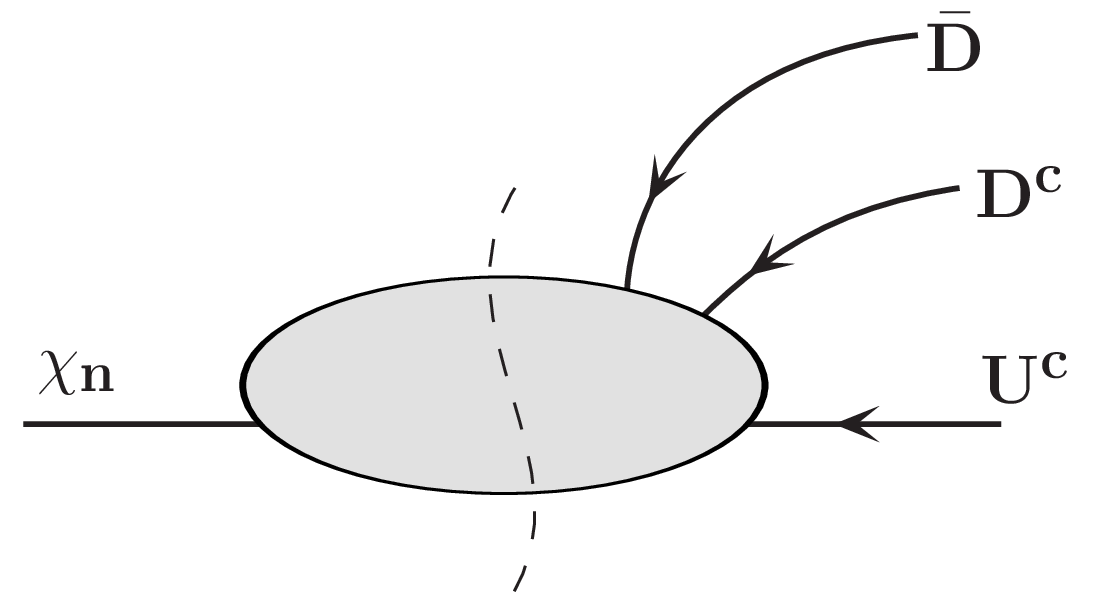}}
  \end{array}  
  \hspace*{-0.25cm} \right|^2
  \] \\
  \vspace*{-0.9cm}
  \beq \hspace*{1cm} A_0 \hspace*{3.25cm} i\hat{A}_1 \hspace*{3.5cm} A^c_0 \hspace*{3cm} i\hat{A}^c_1 \nonumber \eeq
  \vspace*{-0.75cm}
\end{minipage}
\caption{The baryon asymmetry ${\cal A}_B$ from the decay of the Majorana states $\Chi_n$
  is shown as a difference between the rate for the $\Chi_n \to \widebar{D^c}DU$ and the conjugate process
  $\Chi_n \to \widebar{D}D^cU^c$.
  The arrows on the fermion lines show the direction of fermion (baryon) number flow. 
  The dashed curve shows a cut in the loop, contributing a factor of $i$.
  The $a,b,c$ shown next to the momenta are the color indices, and the superscript $c$ on a field denotes its charge conjugate field.
}
\label{ABFeynTh.FIG}
\end{figure}

In Fig.~\ref{ABFeynTh.FIG} we show this situation diagrammatically,
highlighting how the baryon asymmetry arises.
This corresponds to the equation $\AsymB\, \propto\, (\Gamma - \Gamma^{c})\, \propto\, |\ampA_0 + i \hat{\ampA}_1|^2 - |\ampA^c_0 + i \hat{\ampA}^c_1|^2$,
for each $n$,
where the $i$ is due to the cut of the loop diagram shown as the dashed curve in the figure
and is the same for the process and conjugate process, and acts as a strong phase.
The $\hat{\ampA_1}$ itself is complex due to phases in the couplings and we have ${\rm Im}(\hat{\ampA}^c_1) = -{\rm Im}(\hat{\ampA}_1)$, i.e.
these phases act as weak phases. 

\subsubsection{Numerical estimate of $\AsymB$}
\label{ABNumEstm.SEC}

Using Eq.~(\ref{Gm0.EQ}), we make an estimate of the tree-level decay width as
\beq
\Gamma_0 \sim |g|^2\ \left[\frac{M_n^5}{\Lambda^4} \frac{(4\pi)^2}{(2\pi)^5} \hat{f}_{00}\right]  \ ,
\label{Gm0Estm.EQ}
\eeq
where $g$ represents the $G_V$ effective coupling of Eq.~(\ref{LIntVVMB.EQ}), we take $\tilde{g} = 1$ here,
and we introduce the $M_n^5$ for $\Gamma$ to come out with the right dimensions. 
The [...] contains 3-body phase-space factors, namely, 
$(1/(2\pi)^5)$ as usual, a dimensionless $\hat{f}_{00}$ from momentum integration (including Dirac traces), and $(4\pi)^2$ from angular integrals.
For $(G_V) \sim {\cal O}(1)$, we thus obtain the estimate $\Gamma_0/M_n \sim 10^{-3}\, (M_n/\Lambda)^2\, \hat{f}_{00}$.

We estimate next using Eq.~(\ref{Gmhat01.EQ}) the interference term contribution to the decay width as
\beq
\Delta\hat\Gamma_{01} \sim {\rm Im}(g^4) \
 \frac{(4\pi)^2}{(2\pi)^5} \left\{ \frac{M_n^7}{\Lambda^6} \frac{(4\pi)}{(2\pi)^2} ,\ \frac{M_n^9}{\Lambda^8}\, \frac{(4\pi)^2}{(2\pi)^5} \right\} \hat{f}_{01} \ . 
\label{Gm01Estm.EQ}
\eeq
where  
we have shown two possibilities for the cut loop contribution $i\hat\ampA_1$ as \{1-loop,\,2-loop\} factors respectively,    
introducing now the dimensionless $\hat{f}_{01}$ as arising from momentum integrals over phase space and loop momenta (and includes Dirac traces).
The $\{M_n^7,M_n^9\}$ factors is again to obtain the correct dimensions.

Using the estimates in Eqs.~(\ref{Gm0Estm.EQ})~and~(\ref{Gm01Estm.EQ}), from Eq.~(\ref{AsymBhat.EQ}),
we estimate the resulting baryon asymmetry as
\beq
\AsymB \sim \frac{{\rm Im}(g^4)}{|g|^2} \,
 \left\{ \frac{M_n^2}{\Lambda^2} \frac{1}{\pi} ,\ \frac{M_n^4}{\Lambda^4} \frac{1}{2\pi^3} \right\} \frac{\hat{f}_{01}}{\hat{f}_{00}} \ .
\label{AsymBEstm.EQ}
\eeq
For a generic coupling size of about $({\rm Im}(G_V)) \sim 0.1$,
we thus estimate a baryon asymmetry from $\Chi_n$ decay to be of size $\AsymB \sim \{10^{-4}\, M_n^2/\Lambda^2\, ,\ 10^{-5}\, M_n^4/\Lambda^4 \}\, \hat{f}_{01}/\hat{f}_{00}$.
We expect the phase-space and loop-momentum integrals $\hat{f}_{00},\hat{f}_{01} \sim {\cal O}(1)$, but they could be suppressed, for example,
as $(M_U + 2M_D) \to M_n$ closes phase-space, or due to cancellations in angular integrations.
We take up a detailed numerical analysis in Ref.~\cite{OurBGChiFeynNum.BIB} to determine the $\hat{f}_{00},\hat{f}_{01}$,
and compute $\AsymB$ more accurately.

\subsection{Baryon Asymmetry from Scattering}
\label{ABscat.SEC}

Here we discuss the baryon asymmetry in $2\to 2$ scattering processes of the $\Chi_n$ with $Q=\{\Up,\Dp\}$
with cross section $\sigma^n$,
and
compare it with the scattering of the $\Chi_n$ with $\Up^c,\Dp^c$
with cross section $\sigma^{cn}$.
In particular, we consider the scattering processes
\begin{itemize}
\item[] SC-1: $\Chi_n(p_n) \bar\Dp(k_i) \to \Dp(q_1) \Up(q_2)$, and 
\item[] SC-2: $\Chi_n(p_n) \bar\Up(k_i) \to \Dp(q_1) \widebar\Dp^c(q_2)$,
\end{itemize}
for $n=\{1,2\}$. 
SC-1 includes both $\Chi_n(p_n) \Dp^c(k_i) \to \Dp(q_1) \Up(q_2)$ and $\Chi_n(p_n) \bar\Dp(k_i) \to \widebar{\Dp^c}(q_1) \Up(q_2)$.
If the scattering cross sections for process and conjugate-process are different,
i.e. if $\sigma \neq \sigma^c$ for SC-1 and SC-2, we will have a nonzero baryon asymmetry from scattering, $\AsymBsig$. 

The scattering amplitude ${\cal A}^{(\sigma)}$
for each process SC-1 and SC-2
is obtained in close analogy with the decay amplitude of Sec.~\ref{ABdec.SEC}.
This too can be written as a sum over the tree and loop amplitudes,
i.e.,
for the process we write the amplitude as ${\cal A}^{(\sigma)} = {\cal A}_0^{(\sigma)} + {\cal A}_1^{(\sigma)} + ...$,
and for the conjugate process as ${\cal A}^{c\,(\sigma)} = {\cal A}_0^{c\,(\sigma)} + {\cal A}_1^{c\,(\sigma)} + ...$\,.
The scattering cross section for each $n$ is
\beq
\sigma = \frac{1}{v} \frac{1}{2E_n 2E_i} \int [d\Pi_2] \ |\ampA^{(\sigma)}|^2 \ ,
\label{sig.EQ}
\eeq
where
we now integrate over the 2-body final-state phase space,
and
$v$ is the relative velocity between the incoming particles.
The conjugate process cross section $\sigma^c$ is given in terms of $|{\cal A}^{c\,(\sigma)}|^2$.
The tree-level scattering cross section $\sigma_0$ is given by taking $\ampA_0^{(\sigma)}$ in Eq.~(\ref{sig.EQ}).
Since there is no asymmetry at tree level, we have $\sigma_0 = \sigma_0^c$.
Again, like in decay, the scattering baryon asymmetry $\AsymBsig$,
arises to lowest order first in the interference term between the tree and loop amplitudes,
and
the cut in ${\cal A}_1^{(\sigma)}$ adding the piece
for the process $i {\cal \hat{A}}_1^{(\sigma)} \equiv (1/2)\,{\rm Disc}({\cal A}_1^{(\sigma)})$,
and for the conjugate process $i {\cal \hat{A}}_1^{c(\sigma)} \equiv (1/2)\,{\rm Disc}({\cal A}_1^{c(\sigma)})$.
The cut adds to the interference term
the piece $i\hat\ampA_{01}^{(\sigma)} = i{\cal \hat{A}}_1^{(\sigma)}\, {{\cal A}_0^{(\sigma)}}^*$,
and for the conjugate process similarly adds $i\hat\ampA_{01}^{c\,(\sigma)}$.
Correspondingly, this adds to the process cross section $i\hat\sigma_{01}^{(\sigma)}$,
and to the conjugate process it adds $i\hat\sigma_{01}^{c (\sigma)}$.
Analogous to Eqs.~(\ref{AsymBhat.EQ})~and~(\ref{Gmhat01.EQ}),
we now define a cross section asymmetry
\beq
\AsymBsig = \sum_{n=1,2} \AsymBsign \ , \ {\rm with} \ \,
\AsymBsign \equiv \frac{\sigma^{n} - \sigma^{cn}}{\sigma^{n} + \sigma^{cn}}
\approx -\frac{\Delta\hat\sigma_{01}^{n}}{\sigma_0^{n}}
\ , 
\label{ABsigDefn.EQ}
\eeq
where
for leading order accuracy of the asymmetry,
we take $\sigma^{n} \approx \sigma^{cn} \approx \sigma^{n}_0$ in the denominator,
since there is no asymmetry at tree level,
and we now have, for each $n$,
\beq
\Delta\hat\sigma_{01}
= \hat\sigma_{01} - \hat\sigma_{01}^c
= \frac{1}{v} \frac{1}{2E_n 2E_i} \int [d\Pi_2] \ {\rm Im}(\hat\ampA^{(\sigma)}_{01} - \hat\ampA^{c\,(\sigma)}_{01}) = 2\,{\rm Im}(\hat\sigma_{01}) \ .
\label{Dsighat.EQ}
\eeq
This is a way in which a baryon asymmetry might develop due to scattering in our theory,
for which is needed a nonzero ${\rm Im}(\hat\sigma_{01})$.
Diagrammatically, the situation is as shown in Fig.~\ref{ABFeynTh.FIG},
but now with one of the $Q$ legs {\it crossed} from the final state to the initial state as appropriate for SC-1 and SC-2.

\subsubsection{Numerical estimate of $\AsymBsig$}
\label{ABsigNumEstm.SEC}

To estimate $\AsymBsig$ we follow a similar analysis as in Sec.~\ref{ABNumEstm.SEC}, but now, for scattering, take 2-body final state.
In fact, since the scattering matrix element can be got from crossing the corresponding decay amplitude as explained above,
the translation is exact. 
An estimate of the tree-level scattering cross section that follows from taking $\ampA_0^{(\sigma)}$ in Eq.~(\ref{sig.EQ}) is
\beq
\sigma_0 \sim \frac{1}{v} \frac{1}{2E_n\, 2E_i}\, |g|^2 \left[\frac{M_n^4}{\Lambda^4} \frac{4\pi}{(2\pi)^2} \hat{f}_{00}^{(\sigma)} \right]  \ ,
\label{sig0Estm.EQ}
\eeq
where
$g$ represents the $G_V$ effective coupling of Eq.~(\ref{LIntVVMB.EQ}),
we introduce the $M_n^4$ factor for $\sigma_0$ to come out with the right dimensions. 
The [...] factor includes the 2-body phase-space factors, namely, 
$(1/(2\pi)^2)$ as usual, a dimensionless $\hat{f}_{00}^{(\sigma)}$ with Dirac traces (unlike for the 3-body phase space in the decay case, there is no momentum integration for 2-body phase space),
and $4\pi$ from angular integration.
For $g \sim {\cal O}(1)$, $E_n,E_i \sim M_n$, we obtain the estimate $\sigma_0 v \sim 0.1\, (M_n^2/\Lambda^4)\, \hat{f}_{00}^{(\sigma)}$.

We estimate next from Eq.~(\ref{Dsighat.EQ}) the interference term contribution to the cross section as
\beq
\Delta\hat\sigma_{01} \sim \frac{1}{v} \frac{1}{2E_n\, 2E_i}\, {\rm Im}(g^4) \
\frac{4\pi}{(2\pi)^2} \left\{\frac{M_n^6}{\Lambda^6} \frac{4\pi}{(2\pi)^2} \, ,\
                       \frac{M_n^8}{\Lambda^8} \frac{(4\pi)^2}{(2\pi)^5} \right\} \hat{f}_{01}^{(\sigma)}   \ ,
\label{sig01Estm.EQ}
\eeq
where
we have shown two possibilities for the cut loop contribution $i\hat\ampA_1^{(\sigma)}$ as \{1-loop,\,2-loop\} factors respectively, 
and
the $\{M_n^6,\, M_n^8\}$ is again to obtain the correct dimensions for $\Delta\hat\sigma_{01}$.

Using Eqs.~(\ref{sig0Estm.EQ})~and~(\ref{sig01Estm.EQ}) in Eq.~(\ref{ABsigDefn.EQ}),
we estimate the resulting baryon asymmetry as
\beq
\AsymBsig \sim  \frac{{\rm Im}(g^4)}{|g|^2} \,
\left\{ \frac{M_n^2}{\Lambda^2} \frac{1}{\pi} \, ,\ \frac{M_n^4}{\Lambda^4} \frac{1}{2\pi^3} \right\} \frac{\hat{f}_{01}^{(\sigma)}}{\hat{f}_{00}^{(\sigma)}} \ .
\eeq
For a generic coupling size of about ${\rm Im}(g) \sim 0.1$, and $E_n,E_i \sim M_n$, 
we estimate a baryon asymmetry from $\Chi_n$ scattering to be of size $\AsymBsig \sim \{10^{-4}\, M_n^2/\Lambda^2 \, ,\ 10^{-5}\, M_n^4/\Lambda^4 \}\, \hat{f}_{01}^{(\sigma)}/\hat{f}_{00}^{(\sigma)}$.
Comparing with Eq.~(\ref{AsymBEstm.EQ}), we find that the baryon asymmetry from scattering and decay are of similar size.  
We take up a detailed numerical analysis in Ref.~\cite{OurBGChiFeynNum.BIB} to determine the $\hat{f}_{00}^{(\sigma)},\hat{f}_{01}^{(\sigma)}$,
and compute $\AsymBsig$ more accurately.

In Ref.~\cite{OurBGChiFeynNum.BIB} we compute the decay rate and scattering cross section after identifying tree and loop Feynman diagrams
and show that $C$ and $CP$ invariances are violated in these processes, as required to satisfy the Sakharov conditions,
and compute the baryon asymmetry $\AsymB$ and $\AsymBsig$ that are generated.
Next, we place this mechanism of baryon asymmetry generation in a thermal context appropriate for the early Universe
and ask if the observed BAU can be generated via $\Chi$ decays and scattering.

\section{The Baryon Asymmetry of the Universe (BAU)}
\label{BAUgen.SEC}

The baryon asymmetry of the Universe (BAU) today can be expressed as the ratio
\beq
\eta_B \equiv (n_B - n_{\bar{B}})/n_\gamma = 6\times 10^{-10} \frac{\Omega_b h^2}{0.0222} \ ,
\label{etaBDefn.EQ}
\eeq
with $n_B$ ($n_{\bar{B}}$) as the number density of baryons (antibaryons) and $n_\gamma$ as the number density of photons today.
Experimental observations on BBN and CMB~\cite{Planck:2018vyg} tell us that $\eta_B \approx 6\times 10^{-10}$.
In this section, we briefly investigate whether the $\Chi_n$ decay and scattering baryon asymmetry
we identified in Sec.~\ref{ABgen.SEC} could match the observed BAU, and if so,
for what values of parameters of our effective theory.

In the radiation dominated epoch in the early Universe, the $Q=\{\Up,\Dp\}$, being EM charged,
are kept in thermal equilibrium very efficiently by QED processes so that $n_Q = n_Q^{(eq)}$ and $\mu_{\bar Q} = - \mu_Q$,
where $\mu_Q$ ($\mu_{\bar Q}$) is the chemical potential for $Q$ ($\bar{Q}$),
and we write $n_Q = n_Q^{(0)} e^{\mu_Q/T}$, $n_{\bar Q} = n_Q^{(0)} e^{-\mu_Q/T}$. 
The $\Chi$, however, being EM neutral and coupled via the VV interaction, could behave quite differently.  
A nonzero baryon asymmetry could develop if processes involving the $\Chi$ are not fully in thermal equilibrium
owing to $\Chi$ decay and scattering rates $\Gamma_\chi, \Gamma_\chi^{(\sigma)} \lesssim H$, where $H$ is the Hubble expansion rate,
satisfying the Sakharov condition requiring a departure from thermal equilibrium. 
Whether this situation ensues depends on the mass scale $M_\chi$, and the couplings $\tilde{g}, g_{L,R}$.

We analyze this situation by considering the Boltzmann equations (see for example Ref.~\cite{Kolb:1990vq})
that govern the $\Chi,Q,\bar{Q}$ number densities, $n_\chi,n_Q,n_{\bar Q}$ respectively.
The details of setting up these equations for our situation will be presented in Ref.~\cite{OurBGChiCosmo.BIB}, 
and we present here the final result, which is
\bea
\frac{d}{dt}n_\Chi + 3 H n_\Chi\!\!\!\! &=&\!\!\!\! - (\Gamma_\Chi + \Gamma^{(\sigma)}_{\Chi}) (n_\Chi - n_\Chi^{(eq)}) \ , \\
\frac{d}{dt} n_Q + 3 H n_Q \!\!\!\! &=& \!\!\!\! 3 \Gamma n_\Chi - 3 e^{3\mu_Q/T} n_\Chi^{(eq)} \bar\Gamma - (\bar\Gamma^{(\sigma)} - 2 \Gamma^{(\sigma)}) n_\Chi
+ (e^{-\mu_Q/T} \Gamma^{(\sigma)} - 2 e^{\mu_Q/T} \bar\Gamma^{(\sigma)}) n_\Chi^{(eq)} + ... \ , \hspace*{0.75cm} 
%
\label{ChiQBoltzEq.EQ}
\eea
$n_\chi^{(eq)}$ is the equilibrium number density,
the total decay rate $\Gamma_\Chi = \Gamma + \bar\Gamma$ is a sum over the decay rates
$\Gamma = \Gamma(\Chi \to QQQ)$ and $\bar\Gamma = \Gamma^c = \Gamma(\Chi \to \bar{Q}\bar{Q}\bar{Q})$,
their inverse decay channels are
$QQQ \to \Chi$ and $\bar{Q}\bar{Q}\bar{Q} \to \Chi$, respectively,
and all the widths are the thermally averaged widths $\left<\Gamma\right>$. 
The scattering channel cross sections we have used are 
$\sigma \equiv \sigma_{\Chi\bar{Q}} \equiv \sigma(\Chi\bar{Q} \to Q Q)$, 
$\bar\sigma \equiv \sigma^c = \sigma_{\Chi Q} \equiv \sigma(\Chi Q \to \bar{Q} \bar{Q})$, 
their inverse scattering channel cross sections are
$\sigma_{Q Q} \equiv \sigma(Q Q \to \Chi \bar{Q})$,
$\sigma_{\bar{Q}\bar{Q}} \equiv \sigma(\bar{Q} \bar{Q} \to \Chi Q)$, respectively,
and
$\Gamma^{(\sigma)}_{\Chi} = \Gamma^{(\sigma)} + \bar\Gamma^{(\sigma)}$ 
is a sum over the scattering rates 
$\Gamma^{(\sigma)} \equiv \left< \sigma_{\Chi\bar{Q}} v\right> n_{\bar Q}$
and $\bar\Gamma^{(\sigma)} \equiv \left< \sigma_{\Chi Q} v\right> n_Q$,
the $\left< \sigma v\right>$ being the thermally averaged cross section.
We obtain the Boltzmann equation for $n_{\bar{Q}}$ by interchanging
$\Gamma \leftrightarrow \bar\Gamma$, $\Gamma^{(\sigma)} \leftrightarrow \bar\Gamma^{(\sigma)}$,
and $\mu_Q \leftrightarrow -\mu_Q$, in Eq.~(\ref{ChiQBoltzEq.EQ}).
The assumption of $CPT$ invariance implies the following relations among the decay and inverse-decay matrix elements (mod-squared summed over spins),
$|{\cal M}(\Chi \to QQQ)|^2 = |{\cal M}(\bar{Q}\bar{Q}\bar{Q} \to \Chi)|^2$,
and
$|{\cal M}(\Chi \to \bar{Q}\bar{Q}\bar{Q})|^2 = |{\cal M}(QQQ \to \Chi)|^2$, 
and between the inverse and forward scattering channel matrix elements 
$|M|^2_{\bar{Q}\bar{Q}} = |M|^2_{\Chi \bar{Q}}$ and $|M|^2_{Q Q} = |M|^2_{\Chi Q}$.
The baryon asymmetry generation mechanism discussed in Sec.~\ref{ABgen.SEC} leads to
$n_Q \neq n_{\bar Q}$, resulting in the BAU. 
The Boltzmann equation for the net baryon number density $n_B = (n_Q - n_{\bar{Q}})/3$
can now be obtained by taking the difference of the equations for $n_Q$ and $n_{\bar Q}$.
One must correctly match the contributions of an on-shell $\Chi$ intermediate state in other scattering channels,
shown as $...$ in Eq.~(\ref{ChiQBoltzEq.EQ}), with the $\Chi$ decay contribution.
In the small $\mu_Q/T$ limit, Ref.~\cite{Kolb:1979qa} shows that such contributions
overturns the sign of the $\AsymB n_\Chi^{(eq)}$ piece. 
Based on this, we write the Boltzmann equation below with the sign overturned,
and expect that such a mechanism also overturns the sign of the $\AsymBsigHat n_\Chi^{(eq)}$ scattering term.
More details on these aspects will be presented in Ref.~\cite{OurBGChiCosmo.BIB}.
Including these contributions, we write the Boltzmann equation as
\beq 
\frac{d}{dt} n_B + 3 H n_B = \Gamma_\Chi \left[ \AsymB (n_\Chi - n_\Chi^{(eq)}) - n_\Chi^{(eq)} \sinh{\left(\frac{3\mu_Q}{T}\right)} \right] 
+ \Gamma^{(\sigma)}_\Chi \left[\AsymBsigHat (n_\Chi - n_\Chi^{(eq)}) - n_\Chi^{(eq)} \sinh{\left(\frac{\mu_Q}{T}\right)} \right] \ ,
\label{boltzEqnb.EQ}
\eeq
where
we use the definition of $\AsymB$ in Eq.~(\ref{AsymBDefn.EQ}),
we have defined the thermally averaged $\AsymBsigHat = (\Gamma^{(\sigma)}-\bar\Gamma^{(\sigma)})/(2\Gamma_0^{(\sigma)})$ analogous to Eq.~(\ref{ABsigDefn.EQ}),
we have $\sinh{(\mu_Q/T)} = (3/2)\, n_B/n_Q^{(0)}$, 
and we have computed to first order in $Y_B,\AsymB,\AsymBsigHat \ll 1$. 
In Eq.~(\ref{boltzEqnb.EQ}), we see that inverse decay and scattering channels lead to a partial washout of baryon number. 

Changing the independent variable from cosmological time $t$ to the temperature $T$ and writing in terms of $x \equiv M/T$, 
we obtain the equivalent Boltzmann equation for $Y \equiv n/s$ in terms of dimensionless variables as
\bea
\frac{d}{dx} Y_\Chi \!\!\!\! &=& \!\!\!\! - \left(\hat\Gamma_{\rm eff} + \hat\Gamma_{\rm eff}^{(\sigma)}\right) \, x \, (Y_\Chi - Y_\Chi^{(eq)}) \ , \nonumber \\
\frac{d}{dx} Y_B \!\!\!\! &=& \!\!\!\! x \, \hat\Gamma_{\rm eff} \, \left[ \AsymB \, (Y_\Chi - Y_\Chi^{(eq)}) - Y_\Chi^{(eq)} \sinh{\left(\frac{3\mu_Q}{T}\right)} \right]
+ x \, \hat\Gamma^{(\sigma)}_{\rm eff} \, \left[ \AsymBsigHat (Y_\Chi - Y_\Chi^{(eq)}) - Y_\Chi^{(eq)} \sinh{\left(\frac{\mu_Q}{T}\right)} \right]
\,, \hspace*{0.5cm}
\label{BEYchiYB.EQ}
\eea
where
$\hat\Gamma_{\rm eff} = C (M_{Pl}/M_\chi) \hat\Gamma$ with $\hat\Gamma = (\Gamma_\Chi/M_\chi)$,
$\hat\Gamma^{(\sigma)}_{\rm eff} = C' (M_{Pl}/M_\chi) M_\chi^2 \left<\sigma_0 v \right> Y_Q/x^3$,
with $C \sim 1/10$, $C' \sim {\cal O}(1)$,
and 
$\sinh(\mu_Q/T) = (3/2)\, Y_B/Y_Q^{(0)}$. 

The out-of-equilibrium Sakharov condition in our case could be satisfied 
if $Y_\chi$ deviates from $Y_\chi^{(eq)}$ resulting in the BAU.
Equation~(\ref{BEYchiYB.EQ}) indicates that this could happen if
$\hat\Gamma_{\rm eff},\hat\Gamma_{\rm eff}^{(\sigma)} \lesssim 1$. 
From our estimate in Sec.~\ref{ABgen.SEC} that $\hat\Gamma \sim 10^{-5} (M_\chi/\Lambda)^{4}$
and $\AsymB \sim 10^{-5} (M_\chi/\Lambda)^{4}$ (2-loop contribution),
this indicates a mass scale
$M_\chi \gtrsim 10^{13} (M_\chi/\Lambda)^{4}~$GeV.
For instance, for $M/\Lambda \sim 1/10$, the preferred scale is about $10^{9}~$GeV.
These considerations only give an indication of the preferred mass scale in our theory for which the observed BAU is obtained,  
and we will present a more accurate numerical computation of the Boltzmann equations in Ref.~\cite{OurBGChiCosmo.BIB} 
that fully takes into account forward and backward reaction rates in the thermal plasma and possible washout effects,
and distinguishes the $\{\Up,\Dp\}$ in $Q$.

\section{$\Delta B = 2$ Transitions (\nnbar Oscillation)}
\label{DelB2.SEC}

In this section, we discuss $\Delta B = 2$ transitions that violate baryon number by two units,
and focus particularly on the neutron-antineutron \nnbar\!\! oscillation process,
which is a $\Delta B=2$ process since the neutron $n$ has baryon number $B=+1$ and the antineutron $\bar{n}$ has $B=-1$.
We recall that the neutron $n$ at the quark level is the bound state $n=(udd)$ and the antineutron $\bar{n}\equiv (u^c d^c d^c)$, 
which we write symbolically without paying attention to the Lorentz structure but rather emphasizing the internal symmetry quantum numbers,
making no distinction between $\bar{n}$ and $n^c$ (the conjugate).
In our effective theory of Sec.~\ref{Th.SEC}, \nnbar oscillation is induced by the exchange of the Majorana $\Chi_n$ due to its indefinite baryon number.
Since in our VV interaction, two same $\Dp$ fields enter, as discussed in Ref.~\cite{Grojean:2018fus},
the prospects for \nnbar oscillation is better
in comparison with a similar SS operator usually studied in the literature that is forced to have two different flavor $\Dp$ fields.

\nnbar oscillation is being searched for in ongoing experiments, 
and a recent experimental limit on its life-time is $\tau_{n-\bar{n}} \geq 4.7 \times 10^8$~s at 90\% C.L.~\cite{Super-Kamiokande:2020bov}.
The lifetime can be equivalently thought of in terms of a mass difference, namely,
$\Delta m_{n-\bar{n}} \equiv 1/\tau_{n-\bar{n}}$, and the above experimental limit can be recast as
\beq
(\Delta m_{n-\bar{n}})_{\rm expt} \leq 10^{-34}~{\rm GeV} \ .
\label{DmNNBExLim.EQ}
\eeq
In this section, we obtain a rough estimate of the constraint on our effective theory from this limit.

\begin{figure}
  \begin{center}
    \includegraphics[width=0.33\textwidth]{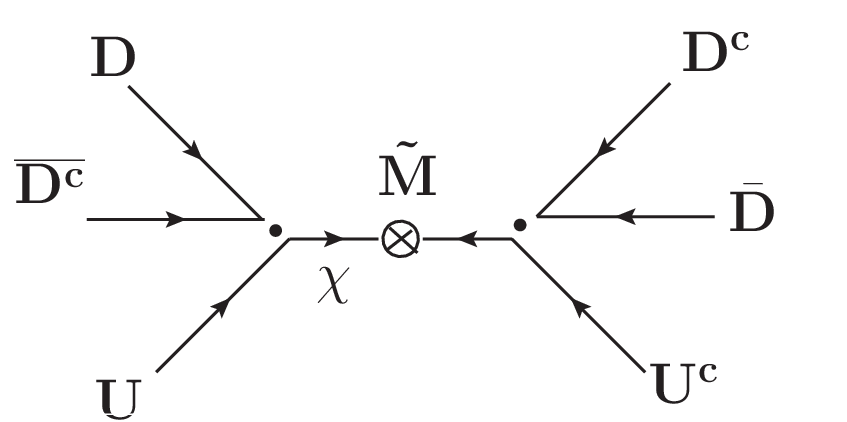}
    \caption{The operator responsible for violation of $B$ by two units that can lead to $N-\bar{N}$ transition.
      \label{NpriNpriB.FIG}
    }
  \end{center}
\end{figure}

A $\Delta B = 2$ process induced in our effective theory of Sec.~\ref{Th.SEC} is shown in Fig.~\ref{NpriNpriB.FIG} where
the $N \equiv (\Dp\Dp\Up)$ transitions to the $\bar{N} \equiv (\Dp^c\Dp^c\Up^c)$,
again symbolically written without paying attention to the Lorentz structure of these operators, 
and making no distinction between $N^c$ and $\bar{N}$.
Integrating out the $\Chi_n$ exchange in Fig.~\ref{NpriNpriB.FIG} generates a six-Q effective operator,
which we encode in an effective Hamiltonian at a renormalization scale $\mu \lesssim M_\chi$ as
\beq
{\cal H}_{\rm eff} \supset {\cal C}_{6Q} {\cal O}_{6Q} = \frac{\tilde{g}^2 \epsilon^{abc} \epsilon^{a'b'c'}}{\Lambda^4 M_\chi} \overline{\Up^c_{c'}} \gamma_\nu \gamma_\mu \Gamma_g \Up_c \ 
\overline{\Dp^c_b} \gamma^\mu \Dp_a \ \overline{\Dp^c_{b'}} \gamma^\nu \Dp_{a'} + {\cal O}(m_n/M_\chi) \ ,
\label{CoOp6Q.EQ}
\eeq
from using the VV interaction couplings of Eq.~(\ref{LIntVVMB.EQ}),
we have set $\Gamma_g \equiv (\ghLn^2\! P_L + \ghRn^2\! P_R)$,
and we have ignored the momentum transfer in the $\Chi$ propagator
since this effective theory is valid for scales $p^2 < M_\chi$.

The discussion in Sec.~\ref{UVcompl.SEC}
on the connection between the $\Up,\Dp$ and the SM quarks $u,d$,
implies the corresponding connection in the effective theory also between the $N$ operator and the neutron $n$,
and similarly between $\bar{N}$ and the antineutron $\bar{n}$. 
In other words, the \NNbar transition induces \nnbar oscillation.
We encode the $Q={\Up,\Dp}$ mixing with the SM $q={u,d}$ in a parameter $s_{\rm eff}$,   
and include in it the $\Dp \leftrightarrow d$, $\Up \leftrightarrow u$ mixing angles $s_{d\Dp}$, $s_{u\Up}$ discussed in Sec.~\ref{UVcompl.SEC}, 
and perhaps CKM-like suppression $s_{\rm CKM}$ if the $\Up$ and $\Dp$ mixing is into 2nd or 3rd generation SM quarks. 
Collecting these factors we write as an example $s_{\rm eff}^2 = s_{d\Dp}^4 s_{u\Up}^2 s_{\rm CKM}^6$,
while reiterating that a different UV realization would have its own structure for the combination of mixing angles. 

We match our $\Delta B=2$ contribution in Eq.~(\ref{CoOp6Q.EQ}) to an effective theory with SM $6q$ operators
listed in Ref.~\cite{Buchoff:2015qwa} (for earlier studies enumerating the operator basis, see references therein). 
To find the overlap of our vector/tensor operator in Eq.~(\ref{CoOp6Q.EQ}) with the standard scalar operator basis
given in Ref.~\cite{Buchoff:2015qwa}, 
we start by denoting our operator as $[\gamma_\nu \gamma_\mu \Gamma_g] (\gamma^\mu) (\!(\gamma^\nu)\!)$
using the bracket convention~\cite{Nishi:2004st} to denote the fields.
We compute the Fierz rearrangement following the method given in Appendix~\ref{Fierz.SEC}, to obtain 
an equivalent Fierz rearranged form of our operator as 
$
- 2 \ghRn^2 [P_R] (P_R)\!) (\!(P_L) - 2 \ghLn^2 [P_L] (P_R)\!) (\!(P_L)
- 2 \ghRn^2 [P_R] (P_L)\!) (\!(P_R) - 2 \ghLn^2 [P_L] (P_L)\!) (\!(P_R)
+ 3 \ghRn^2 [P_R)\!) (P_R] (\!(P_L) - 3 \ghRn^2 [P_R) (P_L)\!) (\!(P_R]
+ 3 \ghLn^2 [P_L)\!) (P_L] (\!(P_R) - 3 \ghLn^2 [P_L) (P_R)\!) (\!(P_L] + ...
$,
where we omit showing the other vector and tensor operators that are generated,
and we have included a minus sign whenever an odd number of fermion fields have been Fierz rearranged. 
These scalar 6-quark operators have overlap with the
$Q_2,\bar{Q}_2$, $Q_3,\bar{Q}_3$, $Q_7,\bar{Q}_7$ operators of Ref.~\cite{Buchoff:2015qwa}. 
An exact matching of our operator with the standard basis including projecting onto the proper tensors in
color space, 
will be taken up in future work~\cite{OurNNbarOsc.BIB}.
Here we content ourselves with estimating the bound on the scale of new physics from the $Q_3$, $Q_7$
contributions we have identified above.
We estimate the \nnbar oscillation rate as
\beq
\Delta m_{n-\bar{n}} \sim \frac{\tilde{g}^2 \GVn \GZn s_{\rm eff}^2}{\Lambda^4 M_n} \matel{Q_i}{\bar{n}}{n}  \ ,
\label{DmnnbarEst.EQ}
\eeq
and include in our estimate the $Q_i = Q_2, Q_3,Q_7$ contributions discussed above.

We use the lattice determination of the matrix elements $\matel{Q_i}{\bar{n}}{n} \equiv \langle Q_i \rangle$
presented in Ref.~\cite{Rinaldi:2019thf},
namely,
$\langle Q_2 \rangle = 144\,(26)$,
$\langle Q_3 \rangle = -47\,(11)$,
$\langle Q_5 \rangle = (-3/2) \langle Q_7 \rangle = -0.23\,(10)$,
in units of $10^{-5}~{\rm GeV}^6$, at the scale $\mu = 700~$TeV.
We can write these matrix elements equivalently as
$\left<Q_i\right> \equiv \Lambda_{\rm QCD}^6 \langle \hat{Q}_i \rangle$ taking $\Lambda_{\rm QCD} = 180~$MeV, 
and from the above values, we have
$\langle \hat{Q}_2 \rangle = 42\,(8)$,
$\langle \hat{Q}_3 \rangle = -13\,(4)$,
$\langle \hat{Q}_7 \rangle = 0.04\,(2)$.  
In our estimate here, we ignore the running from the matching scale $\mu=700~$TeV of Ref.~\cite{Rinaldi:2019thf}
up to the $M_\chi$ scale as it will be a small effect~\cite{Grojean:2018fus}. 

Applying the experimental bound of Eq.~(\ref{DmNNBExLim.EQ}) to the estimate in Eq.~(\ref{DmnnbarEst.EQ}),
we obtain the constraint on the model parameters given by
\beq
\tilde{g}^2 \GVn \GZn s_{\rm eff}^2 \frac{\Lambda_{\rm QCD}^6 \langle \hat{Q}_i \rangle}{\Lambda^4 M_n} \lesssim 10^{-34}\, {\rm GeV} \ ,
\label{nnBOscBnd.EQ}
\eeq
and we use the $\langle\hat{Q}_i\rangle$ given above.
For illustrative purposes, if we take $\tilde{g},\GVn,\GZn \sim {\cal O}(1)$, with $M_n \sim \Lambda$,
we obtain an estimate of the bound on the new physics scale $\Lambda \gtrsim (s_{\rm eff}^{2/5} 10^3)$~TeV
from the $n-\bar{n}$ oscillation experimental constraint.
If $s_{\rm eff}$ is ${\cal O}(1)$, 
the bound we obtain on the scale of new physics is $\Lambda \gtrsim 10^3$~TeV,  
which is well outside the direct on-shell production reach of current (LHC) and planned future colliders,
and only leaves the possibility of probing this physics in precision (indirect) experiments. 
If such new physics is to be accessible kinematically for direct production in present day collider experiments,
we need $\Lambda \sim {\rm TeV}$ for which the bound is $s_{\rm eff} \lesssim 10^{-8}$,
i.e. the neutron mixing to new physics (via the $N$ operator) should be very highly suppressed.
We have already discussed above a natural way this suppression comes about, for instance, due to suppressed mixing angles $s_{\rm qQ}$. 
Taking $s_{d\Dp} \sim s_{u\Up} \equiv s_{\rm qQ}$ we find we need $s_{\rm eff}^{1/3} = s_{\rm qQ} s_{\rm CKM} \lesssim 10^{-3}$ (for $\Lambda \sim $1~TeV),
which is not unreasonable to expect from a UV completion point of view.
As another example, consider the scale obtained in Sec.~\ref{BAUgen.SEC}, namely, $M_\chi \sim 10^9~$GeV, $M/\Lambda \sim 1/10$,  
for which we find $\Delta m_{n-\bar{n}} \sim (\tilde{g}^2 g_{L,R}^2 s_{\rm eff}^2 10^{-18})\, (10^{-34}~{\rm GeV})$. 
Unfortunately, this mass scale is well beyond the reach of current and upcoming \nnbar oscillation experimental searches.
For \nnbar oscillation search prospects in upcoming experiments,
see, for example, Refs.~\cite{Grojean:2018fus,FileviezPerez:2022ypk}.
We leave a more accurate analysis of the \nnbar oscillation rate in our theory with VV interactions for future work~\cite{OurNNbarOsc.BIB}.

\section{Conclusions}
\label{Concl.SEC}

The issue of which physics is responsible for the generation of the observed BAU is presently not yet settled,
although many BSM proposals have been put forth.
In this work we develop an effective theory with
a new Dirac fermion $\chi$ that is uncharged under the SM gauge symmetries but carries nonzero baryon number,
coupled to a $\Up$ and two $\Dp$ fermions, which are up and down type SM-quarklike fermions respectively.
The interaction is a dimension-six effective operator which we denote as $(1/\Lambda^2)(\chi \Up)(\Dp\Dp)$,
not showing Lorentz and color indices, $\Lambda$ being the cutoff scale.
We start by considering both scalar-scalar (SS) and vector-vector (VV) Lorentz structures,
but show that if the two $\Dp$ fermions involved in this coupling are identical, 
the SS interaction is not allowed
owing to the Grassmann nature of the fermion fields and antisymmetry in the color indices,
leaving us to focus on the VV interaction in the remainder of the work.

We speculate on the origin of this effective interaction from a UV completion perspective
and give a few example renormalizable theories,
and comment on other associated operators that are also generated in these examples.
We give an example of how the baryon number violating Majorana mass might arise as a spontaneous symmetry breaking
in a theory which conserves baryon number at the Lagrangian level.
We also give examples of ways the sector in which the baryon asymmetry is generated is connected to the SM.

In our theory, baryon number violation arises due to turning on Majorana masses for the $\chi$,
which splits the Dirac $\chi$ into a pair of Majorana fermions $\Chi_n$ ($n=1,2$) with unequal masses $M_n$.
There are two physical phases in our theory, one in the coupling and another in the Majorana mass.
When we diagonalize the mass matrix and obtain the interaction in the mass basis,
these phases imply complex couplings as shown in Eq.~(\ref{LIntVVMB.EQ}). 
We derive the conditions on the couplings under which the $C$ and $CP$ invariances are broken in our theory,
that along with the source of baryon number violation explained above, 
are required to satisfy the Sakharov conditions.

We consider $\Chi_n$ decay and scattering processes and work out the baryon asymmetry
by comparing the rates for the process and its conjugate process.
We show how the interference between tree- and loop-level amplitudes could result in a baryon asymmetry,
due to the presence of nonzero (weak) phases in couplings that flip sign in going from the process
to its conjugate process,
and a (strong) phase (factor of $i$) coming from on-shell intermediate states in the loop amplitudes
that has the same sign in both.
This is summarized in Fig.~\ref{ABFeynTh.FIG}.

We make numerical estimates for the baryon asymmetry from $\Chi_n$ decays and scatterings,
and find them to be roughly of size $\AsymB, \AsymBsig \sim 10^{-5} \, (M_n/\Lambda)^4$ for reasonable choices of couplings. 
A more detailed numerical analysis considering specific Feynman diagrams is underway, and will be presented
in Ref.~\cite{OurBGChiFeynNum.BIB}.
We place this mechanism of baryon asymmetry generation in the expanding Universe, write down the Boltzmann equation
for the baryon number density and make an estimate of the resulting BAU. 
In another follow-up work~\cite{OurBGChiCosmo.BIB} we will present a more accurate numerical solution of the Boltzmann equation and the BAU.

Another interesting consequence of our theory is the possibility of $\Delta B=2$ transitions,
such as \nnbar oscillation that is being looked for in experiment.
We estimate the size of this oscillation and place constraints on the parameters of the theory.
The $M_\chi$ scale we estimate from requiring that the $\Chi$ be away from equilibrium in order to generate the BAU
suggests that this scale is well beyond the reach of upcoming \nnbar oscillation experiments,
and we will sharpen this estimate in future work~\cite{OurNNbarOsc.BIB}.

More generally, the discovery prospects in ongoing and future experiments depend on the scale of $M_\chi$
and on details of how the communication between the BSM sector and the SM arises.
If $M_\chi \lesssim {\cal O}({\rm TeV})$, there could even be signatures at present day colliders (LHC)
and at proposed hadron colliders, which we relegate to future work.

\medskip
\noindent {\it Dedication:} RT would like to dedicate this paper to the memory of his teacher, Prof. Pranay K. Sen, who passed away on April 17, 2021 due to COVID-19.

\appendix



\section{Spinor algebra}
\label{spinorAlg.SEC}

Here we collect well-known aspects of spinor algebra that are useful to us,
including aspects of 2-component Weyl spinors and its connections to 4-component spinors.
We use the mostly minus $(1,-1,-1,-1)$ metric signature. 
We follow the Van der Waerden undotted and dotted spinor convention to denote spinors that transform as the
$(1/2,0)$ and $(0,1/2)$ irreducible representations of the Lorentz group respectively (cf. Appendix of Ref~\cite{Wess:1992cp}).
 
We write the Lorentz transformation group element as $\mathbb{M} = e^{-(1/2)( i \theta^i + \beta^i) \sigma^i}$,
the $\theta^i, \beta^i$ being 3-rotation and (three) boost parameters respectively, and $\sigma^i$ the three Pauli matrices. 
Under such a Lorentz transformation,
the $\psi_\alpha$ transforms as $\psi_\alpha \to \mathbb{M}_{\alpha\beta} \psi_\beta$ and is said to transform as a $(1/2,0)$ spinor.  
A dual spinor $\chi^\alpha$ can be defined as $\chi^\alpha \equiv \epsilon^{\alpha \beta} \chi_\beta$,
i.e. the index can be raised using the completely antisymmetric tensor ${\boldsymbol{\epsilon}}$,
where we take $\epsilon^{12} = -\epsilon^{21} = +1$.
 Under a Lorentz transformation we have $\psi^\alpha \to \psi^\beta (\mathbb{M}^{-1})^{\beta\alpha}$, 
so that the combination $(\chi\psi) \equiv \chi^\alpha \psi_\alpha$ is a Lorentz invariant
(sum over repeated indices is implied unless specified otherwise).
We denote the inverse of $\epsilon^{\alpha\beta}$ as $\epsilon_{\alpha\beta}$,
and take $\epsilon_{\alpha\beta} = - \epsilon^{\alpha\beta}$ (i.e. $\epsilon_{12} = -\epsilon_{21} = -1$)
so that $\epsilon_{\alpha\beta} \epsilon^{\beta\gamma} = \delta_\alpha^\gamma$.
We then have $\psi_\alpha = \epsilon_{\alpha\beta} \psi^\beta$.
These spinors are taken to be Grassmann objects with $\psi_\alpha \chi_\beta = - \chi_\beta \psi_\alpha$. 
We can show that $(\chi\psi) = (\psi\chi)$. 

We define the conjugate representation, denoted as $\bar\psi_\aldot$, as $\bar\psi_\aldot \equiv (\psi_\alpha)^\dagger$.
Under a Lorentz transformation, it transforms as $\bar\psi_\aldot \to \bar\psi_\bedot (\mathbb{M}^\dagger)_{\bedot\aldot}$
(where $\mathbb{M}^\dagger = {\mathbb{M}^*}^T$ is the adjoint of the matrix $\mathbb{M}$). 
Its dual $\bar\psi^\aldot = (\psi^\alpha)^\dagger$ is given as $\bar\psi^\aldot = \epsilon^{\aldot\bedot} \bar\psi_\bedot$ transforming as 
$\bar\psi^\aldot \to ((\mathbb{M}^\dagger)^{-1})^{\aldot\bedot} \bar\psi^\bedot$ under a Lorentz transformation,
where, $(\mathbb{M}^\dagger)^{-1} = e^{-(1/2)( i \theta - \beta) \cdot \sigma}$.
Similar to the undotted case, we have $\epsilon^{\dot{1}\dot{2}} = -\epsilon^{\dot{2}\dot{1}} = +1$ and its inverse is
$\epsilon_{\dot{1}\dot{2}} = -\epsilon_{\dot{2}\dot{1}} = -1$. 
We can raise and lower similarly, i.e. $\bar\psi^\aldot = \epsilon^{\aldot\bedot} \bar\psi_\bedot$ and $\bar\psi_\aldot = \epsilon_{\aldot\bedot} \bar\psi^\bedot$.
The $\bar\psi^\aldot$ is said to transform as a $(0,1/2)$ spinor.
The combination $(\bar\chi\bar\psi) \equiv \bar\chi_\aldot \bar\psi^\aldot$ is Lorentz invariant, and $(\bar\chi\bar\psi) = (\bar\psi\bar\chi)$. 

The $(\chi\psi)$ is a $c$ number, and its complex conjugate $(\chi\psi)^*$ can be evaluated as 
$(\chi\psi)^* = (\chi^\alpha\psi_\alpha)^* \equiv (\psi_\alpha)^\dagger (\chi^\alpha)^\dagger  = \bar\psi_\aldot \bar\chi^\aldot = (\bar\psi\bar\chi) =
(\bar\chi\bar\psi)$.
We note that the order of the spinors is reversed above when $(...)^\dagger$ is expanded over a spinor bilinear
(without any minus sign).

We find it useful to have a 4-spinor formalism also at our disposal.
We can define a 4-component spinor in terms of the above 2-component (Weyl) spinors as 
\beq
\psi \equiv \bmat \psi_\alpha \\ \bar\chi^\aldot  \emat \ ,
\eeq
where, although we use the same symbol $(\psi)$ for both the 4-component and 2-component spinors,
which one we mean should be clear from the context.
We adhere to the notation of Ref.~\cite{Peskin:1995ev} for the $\gamma^\mu$, namely,
\beq
\gamma^\mu = \bmat 0 & \sigma^\mu \\ \bar\sigma^\mu & 0 \emat \ , 
\eeq
with $\bar\sigma^0 = \sigma^0 = \mathbbm{1}$, $\bar\sigma^i = -\sigma^i$.
We can affix indices as $\sigma^\mu_{\alpha\aldot}$, $\bar\sigma^{\mu\,\aldot\alpha}$, and we have the relation
$\epsilon^{\aldot\bedot} \epsilon^{\alpha\beta}\sigma^\mu_{\beta\bedot} = \bar\sigma^{\mu\,\aldot\alpha}$.
For any 4-component spinor $\psi$, we define the charge-conjugated spinor as $\psi^c \equiv C \psi^* = -i \gamma^2 \psi^*$, 
using the charge conjugation matrix $C=-i\gamma^2$.

Consider a 4-spinor bilinear of the form $B \equiv \overline{\chi} \Gamma \psi$.
We note that the complex conjugate can be evaluated in two ways, namely $B^*$, or as $B^\dagger$.
The first can be written as $\overline{\chi^c} C\,\Gamma^* C\, \psi^c$,
after introducing a minus sign in expanding $*$ over a 4-spinor bilinear.\footnote{
  We note that one must be careful in handling 4-spinor bilinears due to the Grassmann nature of fields. 
  If a matrix notation is followed for 4-spinor bilinears of fields,
   as in usual matrix algebra, we can freely attach a transpose operator $(...)^T$ on a $(1\times 1)$ (i.e. $c$ number) quantity, 
   but, when expanding the transpose over a bilinear of 4-spinor fields, there is an implicit switching of these (Grassmann) fields,
   and therefore a minus sign must be attached.
   Similarly, we can freely attach a dagger operator $(...)^\dagger$ on a $(1\times 1)$ quantity.  
   However, to be consistent with the fact that a $\dagger$ operation expanded over a 4-spinor bilinear switches the order of the
   4-spinors {\em without} a minus sign,
   we also must attach a minus sign when we expand the complex conjugation operator $(...)^*$ over a 4-spinor bilinear.
   Here $(...)^\dagger \equiv (...)^{*\,T}$. 
   These rules on 4-spinor bilinear operations are in place to ensure that the corresponding underlying 2-spinor structure
   is left intact. 
}
The second can be written as
$\overline{\psi} \gamma^0 \Gamma^\dagger \gamma^0 \chi$,
keeping in mind that the $\dagger$ reverses the order of the spinors.
Since for the $c$ number $B$, $B^* = B^\dagger$ what we have shown is 
\beq
\overline{\chi^c} C\,\Gamma^* C\, \psi^c =
\overline{\psi} \gamma^0 \Gamma^\dagger \gamma^0 \chi \ .
\label{chiU4SpRel.EQ}
\eeq
Defining $\Gamma^c \equiv C\,\Gamma^* C$ and $\bar\Gamma \equiv \gamma^0 \Gamma^\dagger \gamma^0$, we can write this as
\beq
\overline{\chi^c} \Gamma^c \psi^c =
\overline{\psi} \bar\Gamma \chi \ .
\label{chiU4SpRelA.EQ}
\eeq
For colored fermions, $\chi_a, \psi_a$, antisymmetrized over the color indices using ${\boldsymbol \epsilon}$,
Eq.~(\ref{chiU4SpRelA.EQ}) can be used to write
\beq
\overline{\chi^c}_a \Gamma^c \, \psi^c_b \, \epsilon^{ab} =
-\overline{\psi}_a \bar\Gamma \chi_b \, \epsilon^{ab} \ , 
\label{chiUCol4SpRel.EQ}
\eeq
where we obtain a minus sign on the RHS from using the antisymmetry property of ${\boldsymbol \epsilon}$. 
We now apply Eq.~(\ref{chiUCol4SpRel.EQ}) to the $\overline{\Dp^c}_a \widetilde\Gamma \Dp_b  \, \epsilon^{ab...} $ part of the interaction of Eq.~(\ref{LeffIntA.EQ})
by taking $\Gamma \to \widetilde\Gamma$, $\chi \to \Dp^c$ and $\psi \to \Dp$, which yields,
\beq
\overline{\Dp}_a \widetilde\Gamma^c \, \Dp^c_b \, \epsilon^{ab} = -\overline{\Dp}_a \widebar{\widetilde\Gamma} \Dp^c_b \, \epsilon^{ab} \ , 
\label{DDspinorRel.EQ}
\eeq
or more simply $\widetilde\Gamma^c = -\widebar{\widetilde\Gamma}$.
This is a nontrivial constraint on $\widetilde\Gamma$. 

Equation~(\ref{DDspinorRel.EQ}) applied to the SS interaction $\LIntSS$, i.e. for $\widetilde\Gamma = \left(\tilde{g}_L P_L + \tilde{g}_R P_R \right)$,
implies that this part of the interaction is the negative of itself, and therefore must be zero! 
This is a consequence of having two identical $\Dp$ fields in $\LInt$, 
the Grassmann nature of the fields, the spinor structure, and the color antisymmetry property. 
Next, Eq.~(\ref{DDspinorRel.EQ}) applied to $\LIntVV$, i.e. for $\widetilde\Gamma = \gamma^\mu \left(\tilde{g}_L P_L + \tilde{g}_R P_R \right)$,
implies that $\tilde{g}_L = \tilde{g}_R \equiv \tilde{g}$ is required for consistency. 

We check the result that the $SS$ interaction cannot be written down now using Weyl spinors.
First note that for a 4-component spinor $\psi = \bmat \psi_\alpha & \bar\chi^\aldot  \emat^T$, we have
$\psi^c = \bmat \chi_\alpha & \bar\psi^\aldot \emat^T$.
Now writing $\Dp = \bmat d_\alpha & {d^c}^\aldot \emat^T$,
where $d_\alpha$ and ${d^c}^\aldot$ are independent Weyl spinors not related by conjugation,
we have $\Dp^c = C \Dp^* = \bmat d^c_\alpha & d^\aldot \emat^T$. 
Then $\overline{\Dp^c}_a (g_L P_L + g_R P_R) \Dp_b  \, \epsilon^{ab...} = (g_L d^\alpha_a {d_\alpha}_b + g_R {d^c_\aldot}_a {d^c}^\aldot_b) \, \epsilon^{ab...} $.
Consider the first term, $d^\alpha_a {d_\alpha}_b \epsilon^{ab} = \epsilon^{\alpha\beta} {d_\beta}_a {d_\alpha}_b \epsilon^{ab}$, 
and since these are Grassmann objects, this is $ - \epsilon^{\alpha\beta} {d_\alpha}_b {d_\beta}_a \epsilon^{ab}$.
Now relabeling $a \leftrightarrow b$ and $\alpha \leftrightarrow \beta$, and using the antisymmetry property of $\boldsymbol{\epsilon}$,
this becomes $-\epsilon^{\alpha\beta} {d_\beta}_a {d_\alpha}_b \epsilon^{ab} = -d^\alpha_a {d_\alpha}_b \epsilon^{ab}$.
In other words we have shown that $d^\alpha_a {d_\alpha}_b \epsilon^{ab}$ is the negative of itself and therefore must be zero.
We can show similarly that ${d^c_\aldot}_a {d^c}^\aldot_b \, \epsilon^{ab} = 0$.
We note that this scalar bilinear is clearly antisymmetic in spin as evidenced by the $\epsilon^{\alpha\beta}$ above. 
Thus we conclude that a scalar (i.e. antisymmetric in spin) Majorana-like bilinear that is
also antisymmetric in color is identically zero. 
We then have shown again, now using Weyl spinors, that the $SS$ interaction cannot be written down.\footnote{We find that
  if we go ahead anyway and compute the tree-level $\Chi_n$ decay amplitude with the SS interaction, we obtain zero,
  consistent with not being able to write down the SS interaction.}

The antisymmetry in color enters crucially in forcing the above Majorana-like scalar bilinear to zero.
This becomes clearer when we consider a similar bilinear, but for a color singlet $\chi$
with 4-spinor $\chi = \bmat \chi_\alpha & {\chi^c}^\aldot  \emat^T$,
i.e., $\overline{\chi^c} (\tilde M_L P_L + \tilde M_R P_R) \chi = (\tilde M_L \chi^\alpha {\chi_\alpha} + \tilde M_R {\chi^c_\aldot} {\chi^c}^\aldot) $.
Going through the same steps as before brings us back to the same form as we started
(without an extra minus sign as in the colored case since we do not have antisymmetric color indices to flip here).
Therefore, this bilinear is not zero and we indeed included this scalar bilinear as the Majorana mass of $\chi$. 

A parity transformation takes $x=(t,\underbar{x}) \to (t,-\underbar{x}) \equiv \tilde{x}$,
under which a 4-spinor transforms as $\psi(x) \to \eta_a \gamma^0 \psi(\tilde{x}) \equiv \widetilde\psi(x)$
(in the notation of Ref.~\cite{Peskin:1995ev} where $\eta_{a,b}$ are introduced), 
and we need $\eta_a = -\eta_b^*$.
For a Majorana fermion $\Chi$ we have $\eta_a = \eta_b \equiv \eta^\Chi$ leading us to take $\eta^\Chi = i$,  
which is also consistent with $\Chi^c = \Chi$.
For the $\Dp,\Up$ fields, we have a choice on what $\eta$ to take, and we find it convenient to make the unconventional choice
$\eta^{\Dp,\Up}_a = \eta^{\Dp,\Up}_b = i$.
We thus have $\widetilde{\psi}(\tilde{x}) = i \gamma^0 \psi(x)$, or equivalently $\psi(x) = -i \gamma^0 \widetilde{\psi}(\tilde{x})$,
and we have $\psi^c(x) = i \gamma^0 \widetilde{\psi}^c(\tilde{x})$, for all the fields $\Dp,\Up,\Chi$.

\subsection{Fierz rearrangement}
\label{Fierz.SEC}
By Fierz rearrangement we can switch the order of
four spinors or fields. Here we start by briefly reviewing the procedure, and then apply it to the VV interaction.

Consider an expression of the form
$\bar{w_1} \Gamma^A w_2 \, \bar{w_3} \Gamma^B w_4$ involving four spinors $w_{1,2,3,4}$ that are $u,v$ spinors,
and $\Gamma^{A,B}$ involve any combination of Dirac matrices.
Using the bracket convention~\cite{Nishi:2004st}, we can denote the above form succinctly as
$(\Gamma^A)[\Gamma^B]$.
The Fierz rearrangement (see, for example, Refs.~\cite{Itzykson:1980rh,Nishi:2004st}) is an equivalent form
written as a linear combination of terms in which the $w_2$ and $w_4$ switch places,
i.e. of the form $\bar{w_1} \hat\Gamma^A w_4 \, \bar{w_3} \hat\Gamma^B w_2$,
which, in the bracket convention, is $(\hat\Gamma^A][\hat\Gamma^B)$.
To derive this form we adopt the chiral basis of Ref.~\cite{Nishi:2004st} and take the 16 matrix basis set as
\beq
\{\Gamma^A\} =\{P_R, P_L, P_R \gamma^\mu, P_L\gamma^\mu, \sigma^{\mu\nu} \} \ ,
\eeq
and the inverse of these as
\beq
\{\Gamma_A\} =\{P_R, P_L, P_L \gamma^\mu, P_R \gamma^\mu, \frac{1}{2} \sigma^{\mu\nu} \} \ .
\eeq
These satisfy the orthogonality condition
\beq
{\rm Tr}(\Gamma_A \Gamma^B) = 2 \delta_A^B \ .
\eeq
The Fierz rearranged combination is given by~\cite{Nishi:2004st}
\beq
(\Gamma^A)[\Gamma^B] = \sum_{C,D} {\cal C}^{AB}_{CD} (\Gamma^D][\Gamma^C) \ ; \quad {\rm where}\ {\cal C}^{AB}_{CD} = \frac{1}{4} {\rm Tr}\left( \Gamma^A \Gamma_C \Gamma^B \Gamma_D \right) \ .
\eeq

Using this procedure, we obtain the Fierz rearrangement of the VV interaction as
%
\beq
(\tilde{g} \gamma^\mu) [\gamma_\mu (g_L P_L + g_R P_R)] = - (\gamma^\mu g_L P_L][\gamma_\mu \tilde{g} P_L) - (\gamma^\mu g_R P_R][\gamma_\mu \tilde{g} P_R) + 2  (g_L P_L][\tilde{g} P_R) + 2  (g_R P_R][\tilde{g} P_L) \ .
\eeq
In terms of the fields, the equivalent Fierz rearranged VV interaction is as shown in Eq.~(\ref{VVFierz.EQ}),
  where we include an extra minus sign arising from switching two fermion fields in the Fierz rearrangement.
%
In particular, we see that we cannot Fierz rearrange our VV operator purely as SS operators,
but rather it will include a sum over other related VV operators (cf. Sec.~\ref{UVcompl.SEC}).

\section{Diagonalization of the $\chi$ sector}
\label{AppChiM.SEC}
To go to the diagonal basis in the $\chi$ sector, we find it useful to write the Dirac (4-component) fermion
$\chi$ in terms of its component Weyl (2-component) fermions
$\chi \equiv \bmat \chi_\alpha & \chi^\aldot  \emat^T$, 
using the dotted and undotted index notation for the Weyl spinors as explained in detail in Appendix~\ref{spinorAlg.SEC}.
In terms of the Weyl spinors, the mass terms in Eq.~(\ref{chiMDir.EQ}) take the form
\beq
    {\cal L}_{\chi\, {\rm mass}} = - \half \bmat \chi^\alpha & {\chi^c}^\alpha \emat
                           \bmat \tilde{M}_L & M_\chi \\ M_\chi & \tilde{M}_R^* \emat
                           \bmat \chi_\alpha \\ \chi^c_\alpha \emat + \hc \ .
\label{LchiM.EQ}                           
\eeq
We recall that $M_\chi$ and $\tilde{M}_L$ are real, while $\tilde{M}_R \equiv \tilde{M}_{R_0} e^{-i\tilde\phi^\prime_R}$ is complex. 

The mass matrix ($M$) in Eq.~(\ref{LchiM.EQ}) is complex symmetric and can be brought to a diagonal form by a rotation of the form
$M_D = \Umat^T M \Umat$, where $\Umat$ is a $2\times 2$ unitary matrix and $M_D$ is real, positive and diagonal. 
We take $\Umat$ to be of the form
\beq
\Umat = \bmat c\, e^{-i \phi_1} & s\, e^{-i\phi_2} \\ -s\, e^{-i (\phi_3 - \phi_2)} & c\, e^{-i(\phi_3 - \phi_1)}  \emat \ ,
\eeq
where $c \equiv \cos{\theta}$ and $s \equiv \sin{\theta}$.
Our task then is to find the $\Umat$, i.e. to find $\theta$, $\phi_{1,2,3}$ that takes us to the diagonal form $M_D$
with real and positive entries.
To reduce clutter, we write $M$ as
\beq
M = \bmat a_0 & b_0 \\ b_0 & d_0 e^{i\phi} \emat \ , \nonumber
\eeq
with 
$a_0 \equiv \tilde{M}_L$, $d_0 \equiv \tilde{M}_{R_0}$, $b_0 \equiv M_\chi$,
$\phi \equiv \tilde\phi^\prime_R$,
and equating the real and imaginary parts of $M_D = \Umat^T M \Umat$ with $M_D$ being real and diagonal, 
we find the solution to be 
\bea
\tan{\tilde\phi} &=& - \frac{d_0 \sin{\phi}}{(a_0 + d_0 \cos{\phi})} \ , \qquad
                       \tan{2 \theta} \equiv t_2 = \frac{2 b_0}{d_0 \cos{(\phi + \tilde\phi)} - a_0 \cos{\tilde\phi} } \ , \\
                       \tan{2\phi_1} &=& \frac{(d_0 s^2 s_{\phi + 2\tilde\phi} - b_0 s_2 s_{\tilde\phi})}{(a_0 c^2 - b_0 s_2 c_{\tilde \phi} + d_0 s^2 c_{\phi + 2\tilde\phi} )} \ , \qquad
                       \tan{2\phi_2} = \frac{(d_0 c^2 s_{\phi + 2\tilde\phi} + b_0 s_2 s_{\tilde\phi})}{(a_0 s^2 + b_0 s_2 c_{\tilde \phi} + d_0 c^2 c_{\phi + 2\tilde\phi} )}  \ , \nonumber \\
                       \phi_3 &=& \phi_1 + \phi_2 - \tilde\phi \ , \nonumber
\eea
written in terms of
$\tilde\phi \equiv \phi_1 + \phi_2 - \phi_3$,
$s_2 \equiv \sin{2\theta}$, $s_{\tilde\phi} \equiv \sin{\tilde\phi}$, $c_{\tilde\phi} \equiv \cos{\tilde\phi}$, 
$s_{\phi + 2\tilde\phi} \equiv \sin{(\phi + 2\tilde\phi)}$, $c_{\phi + 2\tilde\phi} \equiv \cos{(\phi + 2\tilde\phi)}$.

It is possible that a mass eigenvalue (although real) is negative.
If this happens, say for a particular $n$, we multiply the $n$th column of $\Umat$ by $i$, which flips the sign of that eigenvalue
and leads to $M_n$ real and positive.
The new $\Umat$ obtained so by multiplying a column by $i$ remains unitary and is adopted as the $\Umat$ for the rest of the analysis.

With the $\Umat$ as found above, we write the mass eigenstates ${\chi_n}_\alpha = \{ {\chi_1}_\alpha\,\ {\chi_2}_\alpha \}$ as
\beq
\bmat \chi_\alpha \\ \chi^c_\alpha \emat  = \Umat \bmat {\chi_1}_\alpha \\ {\chi_2}_\alpha \emat \ ,
\eeq
which, using an index notation, can be written as ${\chi_a}_\alpha = (\Umat)_{an} {\chi_n}_\alpha$ with
${\chi_a}_\alpha = \{ \chi_\alpha\,\ \chi^c_\alpha \}$ ($a=1,2$ and $n=1,2$). 
We assemble the Weyl mass eigenstate spinors ${\chi_n}_\alpha$ into two 4-component Majorana spinors as
\beq
\Chi_n = \bmat {\chi_n}_\alpha \\ \chi_n^\aldot  \emat \ , 
\eeq
for $n=1,2$.
Clearly, the $\Chi_n$ are self-conjugate and are thus Majorana as claimed. 
We thus have the Dirac 4-component spinor $\chi$ written in terms of two mass eigenstate 4-component Majorana spinors $\Chi_n$ as
\beq
\chi = (\Umat_{1n} P_L + \Umat^*_{2n} P_R) \Chi_n \ .
\label{chi2ChiURot.EQ}
\eeq

The mass eigenvalues, i.e. the entries of $M_D \equiv {\rm diag}(M_1\, , \ M_2)$, are given as 
\bea
M_1 &=& a_0 c^2 \cos{2\phi_1} - b_0 s_2 \cos{(\phi_2 - \phi_1 - \phi_3)} + d_0 s^2 \sin{(\phi-2\phi_3+2\phi_2)} \ ,  \\
M_2 &=& a_0 s^2 \cos{2\phi_2} - b_0 s_2 \cos{(\phi_1 - \phi_2 - \phi_3)} + d_0 c^2 \cos{(\phi-2\phi_3+2\phi_1)} \ . \nonumber
\eea



\end{document}